\documentclass[a4paper, oneside, 11pt]{scrartcl}

\usepackage[utf8x]{inputenc}
\usepackage[T1]{fontenc}

\usepackage[english]{babel}

\makeatletter
\DeclareOldFontCommand{\rm}{\normalfont\rmfamily}{\mathrm}
\DeclareOldFontCommand{\sf}{\normalfont\sffamily}{\mathsf}
\DeclareOldFontCommand{\tt}{\normalfont\ttfamily}{\mathtt}
\DeclareOldFontCommand{\bf}{\normalfont\bfseries}{\mathbf}
\DeclareOldFontCommand{\it}{\normalfont\itshape}{\mathit}
\DeclareOldFontCommand{\sl}{\normalfont\slshape}{\@nomath\sl}
\DeclareOldFontCommand{\sc}{\normalfont\scshape}{\@nomath\sc}
\makeatother

\usepackage[a4paper,  centering, bindingoffset=0mm, 
inner=25mm, outer=25mm, top=26mm, bottom=36mm, heightrounded]{geometry}

\usepackage{graphicx}
\usepackage{float}
\usepackage{cite}

\usepackage{amsmath}	
\usepackage{amssymb}
\usepackage{amsfonts}
\usepackage{mathrsfs}
\usepackage{mathtools}
\usepackage{amsthm}

\usepackage{braket}
\usepackage{slashed}
\usepackage{booktabs}
\usepackage{longtable}	
\usepackage{multirow}

\usepackage{epsf}
\usepackage{epsfig}	
\usepackage{epstopdf}
\usepackage[font=small,labelfont=bf, width=.95\textwidth]{caption}

\usepackage{verbatim}
\usepackage{textcomp}
\usepackage{enumitem}

\usepackage[toc]{appendix}

\usepackage{layout}

\usepackage{microtype}
\usepackage{bookmark}

\usepackage{color}
\usepackage[]{xcolor}

\newcommand{\dd}{\mathrm{d}}

\newcommand{\ii}{\mathrm{i}}

\makeatletter
\newcommand{\subalign}[1]{%
  \vcenter{%
    \Let@ \restore@math@cr \default@tag
    \baselineskip\fontdimen10 \scriptfont\tw@
    \advance\baselineskip\fontdimen12 \scriptfont\tw@
    \lineskip\thr@@\fontdimen8 \scriptfont\thr@@
    \lineskiplimit\lineskip
    \ialign{\hfil$\m@th\scriptstyle##$&$\m@th\scriptstyle{}##$\crcr
      #1\crcr
    }%
  }
}
\makeatother

\usepackage[]{hyperref}

\hypersetup{
    colorlinks=true,%
    citecolor=black,%
    filecolor=black,%
    linkcolor=black,%
    urlcolor=black,
    bookmarksnumbered=true,     
    bookmarksopen=true,         
    bookmarksopenlevel=1,       
    pdfstartview=Fit,           
    pdfpagemode=UseOutlines,
    pdfpagelayout=TwoPageRight
}

\usepackage{graphicx}
\usepackage{tikz}
\usepackage{slashed}
\usepackage{epstopdf}
\usepackage{verbatim}	
\usepackage{blkarray}

\newcommand{\poch}[1]{\left(#1 \right)_n}
\newcommand{\pochm}[1]{\left(#1 \right)}
\newcommand{\sfA}{\mathsf{A}}

\newcommand{\finw}{\text{fin}_w(H_\sfA (\kappa))}

\title{\Huge Feynman integrals as $\sfA$-hypergeometric functions\\[1\baselineskip]}
\author{Leonardo de la Cruz$^a$\\
\emph{\normalfont\normalsize ${}^a$Higgs Centre for Theoretical Physics, School of Physics and Astronomy,}\\
\emph{\normalfont\normalsize The University of Edinburgh,}\\
\emph{\normalfont\normalsize Edinburgh EH9 3FD, Scotland, UK}  
}

\date{%
  $\,$% thats the default I guess
    \\[2\baselineskip]% Space between date and abstract
    \normalfont\normalsize%
    \parbox{0.8\linewidth}{%    
{\bf \sf Abstract}. We show that the Lee-Pomeransky parametric representation of Feynman integrals can be understood
as a solution of a certain  Gel'fand-Kapranov-Zelevinsky (GKZ) system. In order to define such GKZ system, 
we consider the polynomial obtained from the Symanzik polynomials $g=\mathcal{U}+\mathcal{F}$ as having
indeterminate coefficients. Noncompact integration cycles  can be determined from the coamoeba---the argument mapping---of the algebraic variety associated with
$g$. In general, we add a deformation to $g$ in order to define integrals of generic graphs as
linear combinations of their canonical series. We evaluate several Feynman integrals with arbitrary 
non-integer powers in the propagators using the canonical series algorithm. 
    }
}

\begin{document}

\maketitle

\tableofcontents

\section{Introduction}

The analytic evaluation of Feynman integrals in dimensional regularization is still one of the main 
challenges to compute higher order corrections to observables in collider experiments.  Methods for 
evaluating Feynman integrals involve a good understanding of their analytic properties. These have been important from 
the very beginning in order to develop techniques to evaluate them. Long time ago, some of these properties led to the recognition that Feynman integrals  satisfy systems of differential equations
analogous to those of hypergeometric functions\footnote{It is due to Regge the conjecture that Feynman integrals belong to a generalization hypergeometric functions
\cite{de1965differential} and hence that Feynman integrals satisfy analogous systems of differential 
equations. See Ref.\cite{Golubeva1976RuMaS}.}. The modern method to evaluate Feynman integrals is
indeed based on differential equations. It is the combination of Integration by Parts (IBP) identities and differential equations 
\cite{Kotikov:1990kg, Kotikov:1991pm, Remiddi:1997ny, Gehrmann:1999as, Gehrmann:2000zt, Gehrmann:2001ck}.
The result is particularly simple \cite{Henn:2013pwa} if the system  of differential equations evaluates
to combinations of  Multiple Polylogarithms \cite{GONCHAROV1995197, Remiddi:1999ew,Goncharov:1998kja, 
Vollinga:2004sn}.

In favorable cases, Feynman integrals can be evaluated in terms of classical  hypergeometric functions and 
their generalizations such as Appell, Horn, and Lauricella hypergeometric functions. Typically, these functions appear as
infinite sums through the Mellin-Barnes representation of Feynman integrals (see e.g.,\cite{Smirnov:2006ry, Weinzierl:2010ps,
Blumlein:2010zv}).  Once evaluated in terms of hypergeometric functions,  they can be utilized to recover the 
$\epsilon$-expansion \cite{Kalmykov:2006pu, Kalmykov:2006hu, Kalmykov:2007dk, 
Huber:2007dx}, although it is a nontrivial task when hypergeometric functions of many  variables appear
(see e.g., \cite{Moch:2001zr,  Kalmykov:2008ge}).  More recently, the Mellin-Barnes representation has been used to 
find systems of differential equations without resorting to IBP identities \cite{Kalmykov:2012rr},
through the differential reduction method 
\cite{Kalmykov:2006pu, Bytev:2009mn, Bytev:2009kb, Bytev:2011ks,Kalmykov:2011yy, Yost:2011wk, Kalmykov:2012rr, 
Bytev:2013bqa, Bytev:2016ibi}, which is based on Refs. \cite{takayama:1989, takayama:1995}. Using the Mellin-Barnes 
representation it has been shown that a large class of Feynman integrals can be expressed in terms of Horn-type 
functions \cite{Kalmykov:2009tw}. The differential reduction approach has led to the application of techniques coming
from $D$-module theory as has been recently shown in Ref.\cite{Kalmykov:2016lxx}. This point of  
view is close to the one we will adopt in this paper.

$\sfA$-hypergeometric functions were  introduced by Gel'fand-Kapranov-Zelevinsky (GKZ) in 1990 
\cite{GELFAND1990255} as a generalization of the well-known Appell, Lauricella, and Horn series. One important aspect of
the theory is the study of polynomials with indeterminate coefficients associated with an integer matrix $\sfA$\cite{GKZ1995}. This matrix 
and a vector of complex parameters furnish a system of partial differential equations (PDEs) known as a 
GKZ system. Solutions of these systems of PDEs are called $\sfA$-hypergeometric  functions. They can be represented as
 Euler-type integrals and hence these integrals define $\sfA$-hypergeometric functions \cite{GELFAND1990255, 2010arXiv1010.5060N, 2011arXiv1103.6273B, 
2017arXiv170303036F}. On the other hand, series solutions can be computed by a generalization of the Frobenius method 
known as the \emph{canonical series algorithm} due to Saito, Sturmfels, and Takayama \cite{sturmfels:1999}. GKZ systems and 
Feynman integrals have been object of recent interest to mathematicians, who have studied the relation among GKZ systems, Feynman 
integrals, and their regularization \cite{2016arXiv160504970N, Schultka:2018nrs}. Maximal cuts in this language have been 
studied in Ref.\cite{Vanhove:2018mto}.

In this paper, we will consider the parametric representation of Feynman integrals employed by Lee and
Pomeransky to relate critical points of the sum of the Symanzik polynomials and the number of master integrals arising
from IBPs\cite{Lee:2013hzt}. This representation is based on the polynomial $g=\mathcal{U}+\mathcal{F}$,
where $\mathcal{U}$ and $\mathcal{F}$ are the first  and second Symanzik polynomials, respectively. 
We will show that the polynomial $g$ defines a GKZ system, which is constructed 
by considering its coefficients to be indeterminate. We will use this information to obtain series expansions of the 
Euler integrals solutions using the Saito-Sturmfels-Takayama canonical series algorithm. Our definition of a GKZ system based on 
$g$ allows the evaluation of integrals with arbitrary noninteger powers in the propagators using computational algebra. 
The polynomial $g$ may lead to a matrix of codimension $0$. In such cases, we introduce a deformation of
$g$ to ensure a canonical series representation. Once the canonical series are computed and integration
constants are obtained, Feynman integrals can be recovered at the end of the computation by taking the limit of the 
deformation going to zero and setting its coefficients to their kinematic values.

This paper is organized as follows:  In Section \ref{sec2} we review the GKZ framework, present their solutions, and
introduce the canonical series algorithm.  We end this section with examples. In Section \ref{sec3} we introduce 
scalar Feynman integrals and furnish a GKZ system based on them suited for canonical series.  We give examples at the
end of Section \ref{sec3}. We give our conclusions in Section \ref{conclusions}.

\section{$\sfA$-hypergeometric functions and their representations}
\label{sec2}

In this Section, we review basic aspects of the GKZ approach to hypergeometric functions. The main references are the  
book \cite{sturmfels:1999} and the lectures \cite{cattani2006}. Reviews of the main 
concepts can also be found in Refs.\cite{2012Forsgaard, Stienstra:2005nr}. 

This review contains four main parts. We first introduce polynomials with indeterminate coefficients (toric polynomials), which is one of the main ideas
behind the GKZ approach. Polynomials with indeterminate coefficients  lead to generalizations of discriminants, resultants,
and determinants through the study of toric varieties\footnote{An algebraic toric variety is of the form
$\mathbb{C}_*^{n}$, where $\mathbb{C}_{*}=\mathbb{C}\backslash \{0\}$.} \cite{GKZ1995}. 

In the second part, we associate a system of partial differential equations(PDEs) to polynomials with indeterminate
coefficients. We introduce Euler integral representations of solutions of GKZ systems. The system of PDEs may be formally
defined as a  holonomic ideal in a Weyl algebra $D$ and it is the proper language for computational algebra 
purposes \cite{sturmfels:1999}. 

In the third part, we  introduce series representations of solutions of GKZ systems and review the
Saito-Sturmfels-Takayama algorithm to compute canonical series\cite{sturmfels:1999}. The fourth part contains 
examples of the methods introduced. A short example using the computer algebra system \textsl{Macaulay2} can be 
found in Appendix \ref{glossary}.

The connection with Feynman integrals will be made in Section \ref{sec3}. The Saito-Sturmfels-Takayama
algorithm will be our main tool to evaluate Feynman integrals in Sec.\ref{sec3}.

\subsection{Notation}
Throughout this paper we will employ multi-index notation, i.e.,  
\begin{align}
z^\alpha:=&z_1^{\alpha_1} \cdots z_N^{\alpha_N}, \qquad c^{\gamma}:=c_1^{\gamma_1} \cdots c_n^{\gamma_n},\nonumber
\end{align}

\noindent where $\alpha \in \mathbb{K}^N$, $\gamma \in \mathbb{K}^n$. For polynomials $b_1(z), \dots, b_M(z)$ in 
the variables $z=(z_1, \dots,z_N)$, the multi-index notation for products of  polynomials reads 

\begin{align}
b(z)^\beta:=&b_1(z)^{\beta_1} b_2(z)^{\beta_2} \cdots b_M(z)^{\beta_M},\nonumber  
\end{align}

\noindent where $\beta \in \mathbb{K}^M$. Typically we will have $\mathbb{K}=\mathbb{C}$. For $\beta=(1,\dots,1)$, we simply 
write $b(z)^{(1,\dots, 1)}=b(z)$.  In the case $M=1$, we write $b_1(z)^{\beta_1}=b(z)^\beta$. We call polynomials with 
indeterminate coefficients \emph{toric polynomials}\footnote{The reason of this nomenclature is that polynomials with indeterminate 
coefficients will be associated with an integer matrix, which can be interpreted as representing a toric variety.} and emphasize their dependence on their coefficients $c$ by writing $b(c, z)$. The multi-index notation for  differential operators reads $\partial^\alpha=\partial_1^{\alpha_1}, 
\cdots \partial_n^{\alpha_n}$, where $\partial_i=\partial/\partial{c_i}$. Euler operators are defined by

\begin{align}
 \theta_i:=c_i \frac{\partial}{\partial c_i}=c_i \partial_i. \nonumber
\end{align}
\noindent Integrals where toric polynomials are involved will be identified by a subscript, e.g., $I_b$ indicates
that the integral under consideration has a toric polynomial $b(c,z)$ on its integrand. Finally, the 
Pochhammer symbol is defined by 

\begin{align}
\pochm{a}_n:=\frac{\Gamma(a+n)}{\Gamma(a)},\qquad  a \in \mathbb{C}\backslash \mathbb{Z}_{\le 0}.\nonumber     
\end{align}
\noindent Useful identities following from this definition are summarized in Appendix \ref{identities}.

\subsection{Polynomials, varieties, and their coamoebas}
\label{sec:pol-var-coa}
Let us consider $q$ \emph{Laurent} polynomials in  $N$ variables of the form  

\begin{align}
 b_i(z)= \sum\limits_{j=1}^{n_i} c_{ij} z^{\alpha_{ij}}, \qquad c_{{ij}}\in\mathbb{C}_{*}, \qquad i=1,\dots, q, 
\end{align}

\noindent where $\mathbb{C}_{*}=\mathbb{C}\backslash \{0\}$, $\alpha_{ij} \in \mathbb{Z}^N$, and $n_i$ is 
the length of the set of exponent vectors $A_i=\{\alpha_{i1},\cdots, \alpha_{ik}, \cdots, \alpha_{i n_i}\}$ 
associated with the polynomial $b_i(z)$. By abuse of notation, we denote by $A_i$ the $N\times n_i$  \emph{configuration 
matrix} of the exponent vectors of the $i$-th polynomial as

\begin{align}
 A_i=(\alpha_{i1}  \ \cdots \ \alpha_{ik} \ \cdots \alpha_{i n_i}), \qquad \alpha_{i k}\in \mathbb{Z}^N,
\end{align}

\noindent  where each (column) vector  $\alpha_{ik}$ is associated with a monomial term 
$c_{ik}z^{\alpha_{ik}}$ in $b_i(z)$. Therefore $|A_i|=n_i$ is the total number of monomials (columns in $A_i$)\footnote{Here we are adopting a rather unusual route by defining first 
Laurent polynomials and then the matrices $A_i$. The reason behind this is that in Feynman integrals we first consider 
polynomials with determinate coefficients and then consider the indeterminate case. In the mathematics 
literature,  typically one first considers a configuration
matrix and associates a polynomial to it.}. 

Let us give an example. Suppose we have a single polynomial in $N$ variables 
with $n$ monomial terms $b^{\text{ex}}(z)=\sum_{j=1}^n c_j z^{\alpha_j}$, where we have labeled the coefficients simply by
$c_j$, $j=1,\dots,n$. The $N\times n$ configuration matrix of $b^{\text{ex}}(z)$ reads

\begin{align}
A^{\text{ex}}= \begin{blockarray}{cccccc}
c_1 & c_2 & \dots & c_{n-1} & c_n \\
\begin{block}{(c|c|c|c|c)c}
    \alpha_{1,1}& \alpha_{1,2}  & \dots   &  \alpha_{1,{n-1}} & \alpha_{1,{n}} & z_1 \\
   \vdots& \vdots  &   \cdots  &\vdots  &\vdots  &  \vdots \\
   \alpha_{N,1} & \alpha_{N,2} & \dots  & \alpha_{N,{n-1}} & \alpha_{N,n} & z_N \\
\end{block}
\end{blockarray}.
\end{align}
\noindent The monomial term related with the first column reads

$$c_1 z^{\alpha_1}=c_1 z_1^{\alpha_{1,1}} \cdots z_N^{\alpha_{N,1}}.$$

\noindent With the above identifications of the columns and rows, the arrangement of the rows and columns (terms) is irrelevant as they
define the same polynomial. 

Let us now consider the product

\begin{align}
 b(z):= b_1(z) \cdots b_q(z), \label{polynomial-b}
\end{align}

\noindent where each polynomial and its configuration $A_i$ are taken independently, in other words we do 
not expand $b(z)$. Expanding the  polynomials would lead to a single polynomial and hence it is a special 
case of the above. Let $n:=n_1+\dots+n_q$ be the total number of monomials and let us define the  $(N+q)\times n$ matrix

\begin{align}
\mathsf{A}:=
\begin{pmatrix}
\mathsf{1}&\mathsf{0}&\dots& \mathsf{0}\\
\mathsf{0}&\mathsf{1}&\dots& \mathsf{0}\\
\vdots& \vdots&\ddots& \vdots\\
\mathsf{0}&\mathsf{0}& \dots &\mathsf{1}\\
A_1&A_2& \dots&A_q
 \end{pmatrix} \label{matrixA}
\end{align}

\noindent associated with $b(z)$.  Here $\mathsf{0}=(0,\dots,0)$ and  $\mathsf{1}=(1,\dots,1)$ are row 
vectors of length $|A_i|$. We define the codimension of $\mathsf{A}$ as

\begin{align}
 \text{co}(\sfA):= n-N-q.
\end{align}
 
\noindent  The matrix $\sfA$ is the main object of study of the GKZ approach to hypergeometric functions. This 
definition allows us to consider integrals and PDEs later. This matrix is interpreted as representing a 
toric variety\footnote{See Ref.\cite{sturmfels:1996} for an introduction to toric ideals.}.

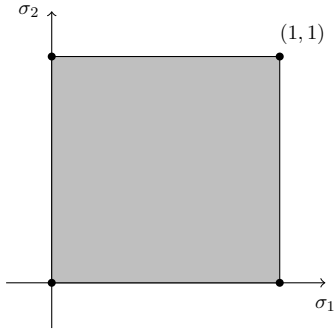
\begin{figure}[ht]
\centering
\begin{tikzpicture}[scale=3]
%\draw[step=1cm,black,very thin] (-1.9,-1.9) grid (1.9,1.9);
\draw[->](-1,-1.2)--(-1,0.2);
\draw[->](-1.2,-1.0)--(0.2,-1);
\draw[fill=black!25] (-1, -1) -- (-1,0) --(0,0) --(0,-1)--cycle; \draw (0,0)
node[fill,circle, inner sep=0pt, minimum size=3pt] {};
\draw (0.1,0.1) node[scale=0.7] {$(1,1)$};
\draw (0,-1) node[fill,circle, inner sep=0pt, minimum size=3pt] {};
\draw (-1,-1) node[fill,circle, inner sep=0pt, minimum size=3pt] {};
\draw (-1,0) node[fill,circle, inner sep=0pt, minimum size=3pt] {};
\draw (-1.1,0.2) node[scale=0.7] {$\sigma_2$};
\draw (0.2,-1.1) node[scale=0.7] {$\sigma_1$};
\end{tikzpicture}
\caption{Polytope of $b(z)=1+z_1+z_2+ c z_1 z_2$.  Written as intersection of half planes, we have 
$\Delta_b= \{\sigma_1\ge 0\}\cap \{\sigma_2\ge 0\}\cap \{-\sigma_1\ge -1\}\cap \{-\sigma_2\ge -1\}$.}
\label{polytope}
\end{figure}

For later purpose we introduce the Newton polytopes of $b_i(z)$ following Ref.\cite{2011arXiv1103.6273B}. 
For each $b_i(z)$, the Newton polytope is the  convex hull $\Delta_{b_i}=\text{conv}(\alpha_{i1}, \dots, 
\alpha_{in_i})$ in $\mathbb{R}^N$. Like any other polytope, we can represent $\Delta_{b_i}$  as the
intersection  of a finite number of halfspaces\footnote{Any polytope has a vertex representation and 
a half space representation. Going from one representation to another involves conversion algorithms
that are beyond the scope of the present work. The interested reader can consult e.g., Ref.\cite{Ziegler:1994}.}:

\begin{align}
 \Delta_{b_i}= \bigcap\limits_{j=1}^{M_i} \{\sigma \in \mathbb{R}^N: 
 \mu_j^i\cdot\sigma\ge \nu_j^i\},
 \label{polytope-half-space}
\end{align}

\noindent where $\mu_j^i \in \mathbb{Z}^N$  are primitive integer vectors in the 
inward normal direction of the facets of $\Delta_{b_i}$, and the $\nu_j^i \in \mathbb{Z}$ are integers 
(See Fig.\ref{polytope}). Here the product $X\cdot Y$ stands for the standard scalar product in $\mathbb{R}^N$.
The polytope of $b(z)$ is the Minkowski sum\footnote{For two polytopes $P$ and $Q$ in 
$\mathbb{R}^N$, their  Minkowski sum $P+Q$ is the set of all vectors $p+q$, such that $p\in P$, $q\in Q$ \cite{Ziegler:1994}.} 

\begin{align}
 \Delta_{b}=  \Delta_{b_1 b_2 \cdots b_q}=  \Delta_{b_1} +\dots + \Delta_{b_q}.
\end{align}

Let us now review the concept of a coamoeba of a variety.  Consider  the ideal $I$ generated by some
polynomials $f_1, \dots, f_j \subset \mathbb{C}[z_1,\dots,z_N]$.  The zero set of the ideal $I=\braket{f_1, \dots, f_j}$ defines the algebraic variety

 \begin{align}
  \mathcal{V}(I) :=\{ z \in \mathbb{C}^N: f_i(z)=0, \; 
  \text{ for all  } f_i \in I \}.
 \end{align}

 \noindent Let us consider the case of  a single polynomial $f$. The  amoeba $\mathcal{A}_f$ of the algebraic
 variety $\mathcal{V}(f)$ is the image of $\mathcal{V}(f)$ under the log mapping
 
 \begin{align}
  \mathcal{A}_f:= \text{Log}(\mathcal{V}(f)),
 \end{align}

 \noindent where $\text{Log}(z)=(\log|z_1|,\dots, \log|z_N|)$.  Amoebas were introduced by GKZ in Ref.\cite{GKZ1995}. 
 Similarly, the coamoeba of $\mathcal{V}(f)$ is the image of $\mathcal{V}(f)$ under the  argument mapping 

\begin{align}
 \mathcal{A}'_f:=  \text{Arg} (\mathcal{V}(f)),
\end{align}

\noindent where $\text{Arg}(z)=(\arg(z_1),\dots, \arg(z_N))$. Here $\arg(z)$ is defined as usual.  For
example, the real positive line is given by $\mathbb{R}_+=\text{Arg}^{-1}(0)$.  

Unlike the log mapping, the argument mapping is multivalued and we can think of 
 it as a multiple periodic subset of $\mathbb{R}^N$ or equivalently it can be viewed as the 
 $N$-dimensional algebraic torus $\mathbb{T}^N=(\mathbb{R}/2\pi \mathbb{Z})^N$ \cite{2011arXiv1103.6273B}. Coamoebas 
 were introduced by Passare in 2004\footnote{Coamoebas have appeared later in the physics literature under the name of algae  in the context of dimer
 models, see e.g.,\cite{Feng:2005gw, 2016arXiv160201826F}.}.
 Our motivation to introduce coamoebas is that they
 define noncompact integration cycles for Euler-Mellin integral representations of $\sfA$-hypergeometric functions.
 These cycles are particularly useful in generic cases where polynomials may vanish on the integration region. In many 
 cases, coamoebas are difficult to study analytically and one  has to consider a rough version, which is 
 called the lopsided coamoeba \cite{Johansson:2010, 2011arXiv1101.4114T, 2012arXiv1205.2014F, 2016arXiv160808663F}. 
 The  relation between $\sfA$-hypergeometric functions and coamoebas has been studied in Refs.\cite{2010arXiv1010.5060N,2016arXiv160808663F, 2011arXiv1103.6273B, 2015Forsgaard}. 
 See Refs.\cite{Johansson:2010, 2012Forsgaard} for an introduction to coamoebas.

\subsection{GKZ systems and $\sfA$-hypergeometric functions}

We denote by  $H_\mathsf{A}(\kappa)$ the GKZ system associated with a matrix $\sfA$ and a 
vector of parameters $\kappa$. It is defined by the following data:

\begin{enumerate}
 \item A  $(N+q) \times n$ matrix $\mathsf{A}$ such that the vector $(1,\dots, 1)$ lies in its row span. 
 Hence, there is a vector $\xi \in \mathbb{Z}^{N+q}$, such that $\xi \mathsf{A}=(1,\cdots ,1)$.
 By definition this matrix is obtained from 
 
 \begin{align}
 b(z)= b_1(z) \cdots b_q(z).
\end{align}

\item A system of partial differential equations(PDEs) associated with $\mathsf{A}$.  Let $u,v \in \mathbb{N}^n$
and consider  

\begin{align}
\Big(\partial^u -\partial^v\Big)F(c)=0&, \quad \text{where} \quad  \mathsf{A}u=\mathsf{A}v,\label{PDE1} \\ 
\left(\sum\limits_{j=1}^n a_{ij} \theta_j-\kappa_i\right)F(c)=0,& \quad 
i=1, \dots, N+q,
\label{PDE2}
\end{align}

\noindent where $a_{ij}$ denotes the components of $\sfA$. Recall that $\theta_j=c_j \partial/\partial{c_j}$ and 
$\partial^u= \partial_1^{u_1} \cdots \partial_n^{u_n}$.

 \item A vector of parameters $\kappa=(\kappa_1,\dots, \kappa_{N+q})$,	 $\kappa_i\in \mathbb{K}$. 
\end{enumerate}
 
 \noindent A holomorphic function $F(c)$  or formal series is called $\sfA$-hypergeometric if it satisfies the above
system of PDEs. 

GKZ systems can be rigorously defined in the language of holonomic ideals in the ring
of differential operators with polynomial coefficients---the so called  Weyl algebra 
$D=\mathbb{K}\braket{c_1,\dots ,c_n,\partial_1, \dots,\partial_n}$ modulo commutation rules. In this sense, the 
\emph{toric ideal} associated with $\sfA$ is defined by 

\begin{align}
 I_\sfA:=\braket{\partial^u-\partial^v: \sfA u=\sfA v, \quad u,v\in \mathbb{N}^n} 
 \subset \mathbb{K}[\partial_1,\dots,\partial_n],
 \label{IA-toric}
\end{align}

\noindent where   $\mathbb{K}[\partial_1,\dots,\partial_n]$ is a commutative polynomial ring. In addition, we construct
the ideal generated by the column vectors $\kappa^T$ and $\theta=(\theta_1, \dots, \theta_n)^T$. This ideal is given 
by 

\begin{align}
 \braket{\sfA \theta-\kappa^T}\subset \mathbb{K}[\theta_1,\dots,\theta_n],
\end{align}

\noindent where each generator of the ideal has the form $\sum_{j=1}^n a_{ij} \theta_j-\kappa_i$. 
The GKZ system $H_{\sfA}(\kappa)$ denotes the left ideal on the Weyl
algebra $D$ generated by $I_\sfA$ and $\braket{\sfA \theta-\kappa^T}$.  In this language a holomorphic function $F(c)$
or formal series is called $\sfA$-hypergeometric of degree $\kappa$ if 
$H_{\sfA}(\kappa) \bullet F(c)=0$, where $\bullet$ denotes the action of the Weyl algebra on polynomials
\cite{sturmfels:1999}. The language of holonomic ideals and $D$-modules is the appropriate one to treat the 
problem using computational algebra.

Let us denote by $\text{vol}(\mathsf{A})$ the normalized volume---w.r.t the volume of the standard simplex
which is equal to 1---of the convex hull of $\sfA$\footnote{For the case $q=1$, this normalization implies 
that $\text{vol}(\sfA)=N! \text{vol}(\Delta_b)$ with $\Delta_b$ defined in Sec.\ref{sec:pol-var-coa}.}. Then, for 
generic parameters\footnote{Here, generic parameters $\kappa\in \mathbb{K}^{N+q}$ are to be understood as ranging over 
nonempty open algebraic (Zariski) subsets of $\mathbb{K}^{N+q}$.} $\kappa$, the rank of the system satisfies the inequality 
(Theorem 3.5.1 in \cite{sturmfels:1999})

\begin{align}
 \text{rank}(H_\mathsf{A}(\kappa)) \ge \text{vol}(\mathsf{A}), 
\end{align}

\noindent which corresponds to the dimension of the solution space. This number can also be computed 
from the degree of the toric ideal $I_\sfA$.

For generic parameters $ \text{rank}(H_\mathsf{A}(\kappa))=\text{vol}(\mathsf{A})$. In cases where the parameters have certain special
values (nongeneric) the rank of the system can jump and we have  $\text{rank}(H_\mathsf{A}(\kappa)) > \text{vol}(\mathsf{A})$
(Example 4.2.7 in Ref.\cite{sturmfels:1999}.).

\subsubsection{Euler-type integral solutions}
\label{Euler-type-integrals}
Solutions of GKZ systems  have representations as Euler-type integrals. Accordingly, we call them 
$\sfA$-hypergeometric functions. They can be constructed by taking the vector of parameters

\begin{align}
 \kappa=(-\beta, -\alpha), \qquad \beta \in \mathbb{C}^{q}, \quad \alpha \in\mathbb{C}^N, 
\end{align}

\noindent and then write the integral associated with $\sfA$ as follows:

\begin{align}
 I_b (\kappa)= \int_{\Omega}  \frac{z^\alpha}{b(c, z)^\beta} \dd\eta_N , \qquad 
 \dd \eta_N= \frac{dz_1}{z_1} \wedge \frac{dz_2}{z_2} \wedge \cdots  \wedge \frac{dz_N}{z_N},
 \label{EM-rep}
\end{align}

\noindent where the integration cycle is such that $\Omega \subset (\mathbb{C}_{*})^N\backslash 
\mathcal{V}(b)$. It is usually assumed that the cycles are compact \cite{GELFAND1990255}.   

In Ref.\cite{2011arXiv1103.6273B}, Berkesh, Forsg{\aa}rd, and Passare (BFP) constructed explicit noncompact 
cycles for these type integrals. In order to reach the above type of integral, BFP proceed in three steps.  
\emph{First}, we consider  that $b(z)$  does not vanish in the positive 
orthant and consider the \emph{Euler-Mellin} integral 

\begin{align}
 I(\kappa)= \int_{\mathbb{R}_+^N}  \frac{z^\alpha}{b(z)^\beta}  \dd\eta_N=
 \int_{\mathbb{R}^N} \frac{e^{(\alpha, x)}}{b(e^x)^\beta} \dd x,\qquad \dd x= d x_1 \wedge \cdots  \wedge d x_N,
 \label{EM-BFP}
\end{align}

\noindent where  in this case  polynomial coefficients are fixed. BFP showed that 
if the polynomials $b(z)$ are nonvanishing\footnote{Technically, these polynomials should 
be completely nonvanishing in the sense of BFP i.e., not vanishing on the faces of the 
polytope of $b(z)$. See definition 2.1 of Ref. \cite{2011arXiv1103.6273B}. 
See also \cite{2012Forsgaard}.}, then   Eq.\eqref{EM-BFP} converges and defines and analytic
funcion with parameters $\kappa=(-\beta, -\alpha)$ on the tube domain 

\begin{align}
\{(\alpha, \beta) \in \mathbb{C}^{N+q} | \tau :=\text{Re } \beta\in \mathbb{R}_+^q, \qquad \sigma:=\text{Re } \alpha 
\in \text{int}(\tau \Delta_b ) \}, 
\end{align}

\noindent where $\text{int}(\tau \Delta_b)$ is the interior of  of the weighted Minkowski sum of the Newton 
polytopes of $b_j$ weighted by $\tau$, i.e., $\tau \Delta_b=\sum_{j=1}^q \tau_j \Delta_{b_j}$. The
\emph{second} step is to consider the less favorable case where the polynomials $b(z)$ vanish on the 
positive orthant. Here, we can take a  connected component $\Theta$  
of  $\mathbb{R}^N\backslash \overline{\mathcal{A}}'_b$, where $\overline{\mathcal{A}}'_b$ denotes the closure of 
the coamoeba of $b$  and consider the integral

\begin{align}
 I(\kappa)= \int_{\text{Arg}^{-1}\theta}  \frac{z^\alpha}{b(z)^\beta} \dd\eta_N  =	
 \int_{\mathbb{R}^N}  \frac{e^{\alpha\cdot(x+\ii \theta)}}{b(e^{x+\ii \theta})^\beta}  \dd x , 
 \label{integral-coamoeba}
\end{align}

\noindent where  $\theta \in \Theta $ is a representative of the connected component of complement of 
coamoeba of $b(z)$. This is essentially  a change of variables and a
slight perturbation of $\theta$ does not impact the result of the
integral. We write an analogue of Theorem 2.4  in Ref.\cite{2011arXiv1103.6273B}, where the proof can be 
found.

\subsubsection*{Theorem(Berkesh, Forsg{\aa}rd, and Passare)}

\emph{For nonvanishing polynomials, $b_1, \dots,b_q$ in $\text{Arg}^{-1}\theta$  the 
integral \eqref{integral-coamoeba} admits a meromorphic continuation of the form
}
\begin{align}
 I(\kappa)= \Phi_b^{\Theta}(\alpha,\beta) \prod\limits_{k=1}^M
 \Gamma(\mu_k\cdot \alpha- \nu_k\cdot \beta),  
\end{align}

\noindent \emph{where $\Phi_b^{\Theta}(\alpha,\beta)$ is an entire function and $\Theta$ is a connected component. 
$\mu_k$, $\nu_k$ can be recovered from the Newton polytope $\Delta_{b_1\cdots b_q}$ (See Eq.\eqref{polytope-half-space}).}

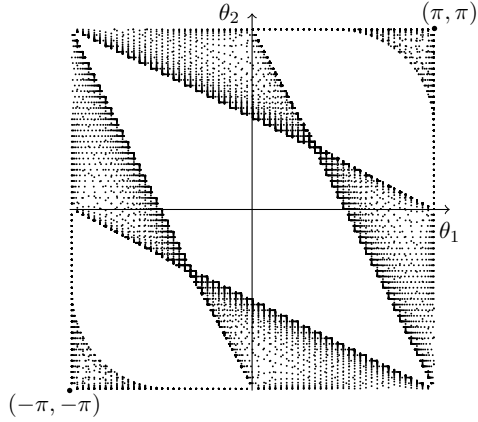
\begin{figure}[ht]
 \centering
 \begin{tikzpicture}
\node[anchor=south west,inner sep=0] at (0,0)
{\includegraphics[width=0.3\textwidth]{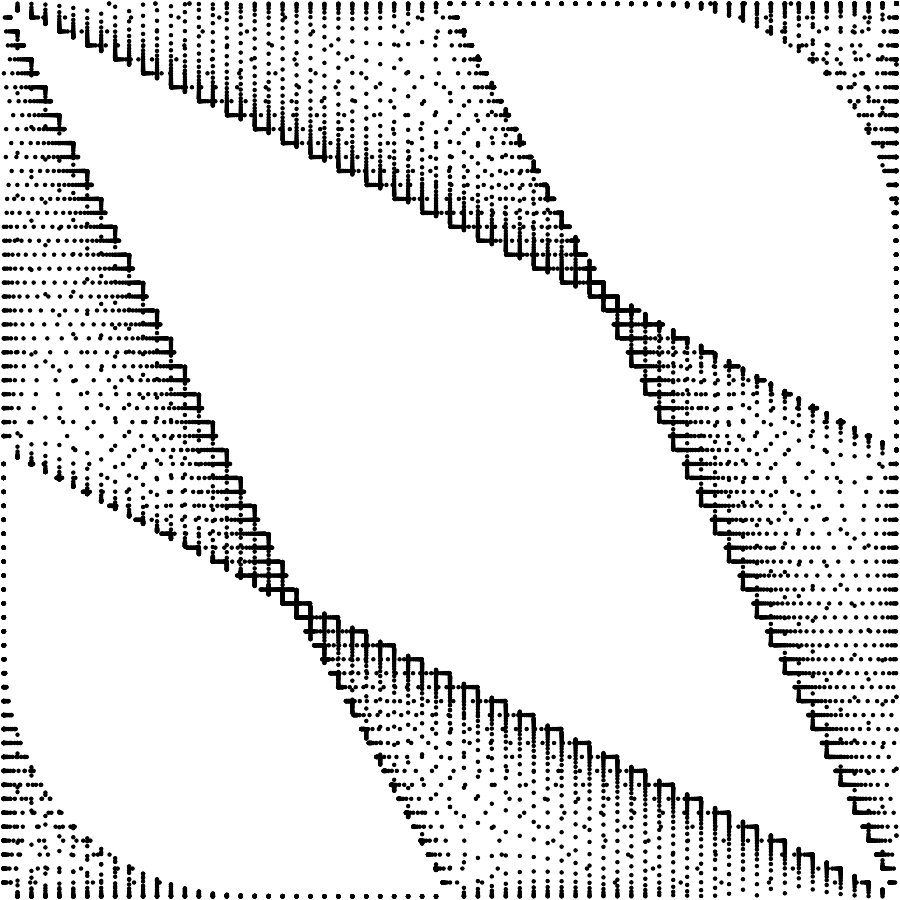}};
\draw [->](0,2.4) -- (5.0,2.4);
\draw [->](2.4,0) -- (2.4,5.0);
\node[scale=0.8, text centered] at (-0.2,-0.2) {$(-\pi, -\pi)$};
\node[scale=0.8, text centered] at (5,5) {$(\pi,\pi)$};
\node[scale=0.8, text centered] at (2.1, 5) {$\theta_2$};
\node[scale=0.8,text centered] at (5,2.1) {$\theta_1$}; 
\draw (4.8,4.8) node[fill,circle, inner sep=0pt, minimum size=2pt] {};
\draw (0,0) node[fill,circle, inner sep=0pt, minimum size=2pt]{};
\end{tikzpicture}
 \caption{Coamoeba (shaded) of $b=1+z_1z_2 +z_1 z_2^2+z_1^2  z_2^2 +   c_4 z_1^2 z_2 $ drawn in the fundamental
domain $[-\pi, \pi] \times  [-\pi, \pi]$ of $\mathbb{T}^2$ in $\mathbb{R}^2$. This case corresponds
to $c=(1,1,1,1,c_4)$, with $c_4$ near $1$. Here, we have three connected components, one of which contains $(0,0)$ (center of the Figure).}
 \label{coamoeba-curve}
\end{figure}

The \emph{third} step is the transition to $\sfA$-hypergeometric functions by promoting the coefficients of 
$b(z)$ in Eq.\eqref{integral-coamoeba} to indeterminate, therefore we consider them as variables. 
The integral\footnote{Here $c\in \mathbb{C}^n \backslash \Sigma_\sfA$, where $\Sigma_\sfA$ is the singular 
locus of all $\sfA$-hypergeometric functions.}

\begin{align}
I_b(\kappa)=
 \int_{\text{Arg}^{-1}\theta}  \frac{z^\alpha}{b(c, z)^\beta} \dd\eta_N
 \label{integral-representation-coamoeba}
\end{align}
\noindent is a representation of an $\sfA$-hypergeometric function (Theorem 4.2 in 
\cite{2011arXiv1103.6273B}).  For generic parameters $\kappa$,  Eq.\eqref{integral-representation-coamoeba} provides a basis of solutions
of $H_{\mathsf{A}}(\kappa)$,  where each integral is  evaluated on a representative  of $\Theta$ for each
connected component of $\mathbb{R}^N\backslash  \overline{\mathcal{A}}'_b$ 
(See Fig.\ref{coamoeba-curve}\footnote{We thank Jens 
Forsg{\aa}rd for providing his \textsc{Mathematica} package to draw coamoebas and lopsided coamoebas.} ).

\subsection{$\sfA$-hypergeometric canonical series}

Saito, Sturmfels, and Takayama (SST) generalized the Frobenius method using Gr\"obner deformations in order to deal with
regular holonomic systems (Chapters 2 and 3 of \cite{sturmfels:1999}). Roughly speaking, it consists on taking certain initial
ideals of $I_\sfA$ (Eq.\eqref{IA-toric}) with respect to a weight $w\in \mathbb{R}^n$ and generate the series 
solutions from it. The role of the \emph{indicial equation} in the Frobenius method
is played by an \emph{indicial ideal} $\text{ind}_w(I_{\sfA})$ with respect to $w$, together with the 
ideal $\braket{\sfA\theta-\kappa^T}$. As in the Frobenius method, 
the problem consist of finding the roots $\gamma$ of those ideals and generate the coefficients
of the series. Series thus obtained belong to the Nilsson
ring, i.e, series of the form 

\begin{align}
 f= \sum\limits_{\alpha,\beta}  k_{\alpha\beta}\; c^\alpha \log (c)^{\beta}. 
\end{align}

In this paper, we will be chiefly interested in the case where the resulting series are 
\emph{logarithm-free}, i.e., where $\beta=0$ in the above equation. Let us turn our attention to these cases. Let 

\begin{equation}
\mathcal{L}:=\{u\in \mathbb{Z}^n : \sfA u=0\} 
\end{equation}

\noindent be a lattice of rank $m$ and let $\kappa^T =\sfA \gamma^T$ \footnote{Notice that we will not be interested
in general vectors $\gamma \in \mathbb{C}^n$ but only those which are the roots of an ideal of $H_\sfA(\kappa)$ with 
respect to some weight vector $w$. See Algorithm below.}. For 
$u\in \mathcal{L}$ we can write $u=u_+-u_-$, where $u_{\pm} \in \mathbb{N}^n$ have disjoint support. Now, 
for $\gamma \in \mathbb{C}^n$  we define  the following  quantities expressed as falling factorials 

\begin{align}
 [\gamma]_{u_-}:=& \prod\limits_{i:u_i<0}\;\; \prod\limits_{j=1}^{-u_i}(\gamma_i-j+1)=
 \prod\limits_{i:u_i<0}(-1)^{-u_i} \pochm{\gamma_i}_{-u_i},\\
  [\gamma+u]_{u_+}:=& \prod\limits_{i:u_i>0} \;\; \prod\limits_{j=1}^{u_i}(\gamma_i+u_i-j+1)=
  \prod\limits_{i:u_i>0} \prod\limits_{j=1}^{u_i}(\gamma_i+j)=
  \prod\limits_{i:u_i>0}\pochm{\gamma_i+1}_{u_i},
\end{align}

\noindent where $(a)_x$ are Pochhammer symbols. Then for $\gamma \in \mathbb{C}^n$, such 
that no element in $\gamma$ is a nonnegative integer, the series  

\begin{align}
 \phi_\gamma:= \sum\limits_{u \in \mathcal{L}} \frac{[\gamma]_{u_-}}{[\gamma+u]_{u_+}} c^{(\gamma+u)} 
 \label{log-free-series}
\end{align}

\noindent  is a solution of $H_\mathsf{A}(\kappa)$ (See proposition 3.4.1 and Theorem 3.4.2 in Ref.\cite{sturmfels:1999}).
When the vectors $\gamma$ contain negative integers, we define the negative support of $\gamma$ as

\begin{align}
\overline{\text{supp}}(\gamma)=\{i \in\{1,\dots, n\}: \gamma_i \in \mathbb{Z}_{<0}\}.  
\end{align}

\noindent In those cases, we  consider the lattice 
\begin{align}
 \mathcal{N}_\gamma:=\{u\in \mathcal{L}| \overline{\text{supp}}(\gamma+u)= 
\overline{\text{supp}}(\gamma)\}
\end{align}
\noindent and perform the sum over $\mathcal{N}_\gamma$ in Eq.\eqref{log-free-series}. In this
paper, we will assume that the vectors $\kappa$ are generic and thus none of the roots $\gamma$ are negative integers\footnote{Corollary 3.4.3 in
Ref.\cite{sturmfels:1999}.}.
This implies that $\overline{\text{supp}}(\gamma)=\emptyset$ and  therefore the sum runs over $u \in \mathcal{L}$.

The roots $\gamma$ can be obtained by finding the roots of an ideal known  as \emph{fake indicial ideal}. 
Accordingly, the roots are called \emph{fake exponents} and we give the algorithm to compute them below. 
Series thus obtained are called \emph{canonical series}.  The series obtained from this algorithm
have a common domain of convergence $\mathcal{U}_w\in \mathbb{C}^n$, which is characterized by a weight vector $w$.  For
generic roots $\gamma$, canonical series provide a basis of holomorphic solutions of  $H_\sfA(\kappa)$ in
$\mathcal{U}_w$ (For a description of this domain see Theorem 2.5.16 in Ref.\cite{sturmfels:1999} and
Theorem 3.19 in Ref.\cite{cattani2006}). 

The relation with Euler integrals goes as follows. Suppose we find $r$ different 
roots. The general solution of the integral \eqref{EM-rep} is given by a linear combination of its canonical
series \cite{sturmfels:1999}, i.e., 

\begin{align}
I_b(\kappa)= K_1\phi_{1} +\dots +K_{r} \phi_{r},  
\label{linear-combination}
\end{align}

\noindent where the canonical series $\phi_r$ can be computed  independently of the cycle $\Omega$. The integration cycle plays a 
role in the computation of the integration constants $K_i$. We discuss how to obtain integrations constants for the particular case of 
Feynman integrals in the next section.

We proceed with the canonical series algorithm to compute fake exponents $\gamma$.  A comprehensive review  of 
this algorithm  can be found in Ref.\cite{cattani2006}. We start with some definitions.

\subsubsection*{Definitions}
\emph{Toric ideal.} Let $D$ be a Weyl algebra over $\mathbb{K}$, which is a free (noncommutative) 
associative $\mathbb{K}$-algebra generated by $\mathbb{K}\braket{c_1,\dots, c_n, \partial_1,\dots,\partial_n}$ modulo
the commutation rules
\begin{align}
c_ic_j=c_jc_i, \qquad \partial_i \partial_j = \partial_j\partial_i, \qquad  \partial_i c_j=  \delta_{ij}.
\end{align}
\noindent Let $\sfA$ be as in Eq.\eqref{matrixA}, then 
\begin{align}
 I_\sfA:=\braket{\partial^u-\partial^v: \sfA u= \sfA v, \qquad  u,v \in \mathbb{N}^n}
\end{align}

\noindent generates the \emph{toric ideal} associated with $\sfA$. Toric ideals can be computed 
using reduced Gr\"obner bases\cite{sturmfels:2001}.

\emph{Initial ideal.} Let $w\in \mathbb{R}^n$ be a weight vector and let $I_\sfA$ be a toric ideal. 
For $w$ generic, the ideal $\text{in}_w(I_A)$ is a monomial ideal generated by the leading terms of $I_A$ 
with respect to the partial ordering $\preceq_w$. 

\emph{Standard pairs.} Let $R=\mathbb{K}[\partial_1,\dots, \partial_n]$ and let $I$ be a  monomial ideal in $R$. 
Furthermore, let  $\partial^\alpha$ be a monomial and $F\subseteq \{1, \dots,n\}$, where $\alpha\in \mathbb{N}^n$.
A standard pair of a monomial ideal $I$ is a pair $(\partial^\alpha, F)$ satisfying 
three conditions:

\begin{enumerate}
 \item $\alpha_i= 0$ for all $i\in F$,
 \item for all choices of integers $\beta_j\ge 0$, the monomial $\partial^\alpha\prod_{j\in F} \partial_j^{\beta_j}\notin I$,
 \item for all $l\notin F$, there exist $\beta_j\ge 0$ such that $\partial^\alpha\partial_l^{\beta_l}\prod_{j\in F} 
 \partial_j^{\beta_j} \in I$.
\end{enumerate}
\noindent Let us denoted by $\mathcal{S}(I)$  the set of all standard pairs of $I$. The decomposition of $I$ into 
irreducible monomial ideals can be obtained from the identity.
\begin{align}
 I= \bigcap\limits_{(\partial^\alpha, F) \in \mathcal{S}(I)} \braket{\partial_i^{\alpha_i+1}:i\in F} .
\end{align}

\subsubsection{Algorithm (Saito-Sturmfels-Takayama)}
\label{SST-algorithm}

In this algorithm we will set $\mathbb{K}=\mathbb{C}$.\\

\noindent Input: Matrix $\sfA$, weight vector $w$, and complex parameters $\kappa$.  

\noindent Output: Roots of the fake indicial ideal $\finw$.

\begin{enumerate}
 \item Compute the toric ideal associated with $\mathsf{A}$
 
 \begin{align}
  I_\mathsf{A}=\braket{\partial^u-\partial^v:\mathsf{A} u= \mathsf{A}v, \qquad  u,v \in \mathbb{N}^n}.
 \end{align}
 
\noindent Notice that this is an ideal in the commutative polynomial  ring 
$\mathbb{C}[\partial_1,\dots,\partial_n]$. This ideal is the input to compute the combinatorial object of 
standard pairs  $\mathcal{S}$ and the initial ideal with respect to $w$.   

\item Let $w \in\mathbb{R}^n$ be a generic weight vector. Compute the initial ideal $\text{in}_w(I_\mathsf{A})$ with 
respect to $w$  and obtain its standard pairs $\mathcal{S}( \text{in}_w(I_A))$. Standard pairs
are combinatorial objects which tells us the types of solutions.

\item Use the standard pairs to construct the  indicial ideal

\begin{align}
\text{ind}_{w}(I_\sfA)= 
\bigcap\limits_{(\partial^a, F)\in \mathcal{S}( \text{in}_w(I_\sfA))}
\braket{(\theta_j-a_j), j\notin F} \subset \mathbb{C}[\theta_1, \theta_2, \dots,\theta_n],
\end{align}
\noindent where $\theta_i=c_i \partial_i$. 

\item Write the ideal $\braket{\sfA \theta -\kappa^T} \subset \mathbb{C}[\theta_1, \theta_2, \dots,\theta_n] $.

\item The fake indicial ideal with respect to $w$ is given by 

\begin{align}
\finw:=& \text{ind}_{w}(I_\sfA)+\braket{A\theta-\kappa^T}.
\end{align}

\item Compute the roots of $\finw$. These are called fake exponents and we denote them by $\gamma$.
\end{enumerate}

The canonical series are then given by Eq.\eqref{log-free-series}. In order to write solutions, we compute the kernel 
of $\mathsf{A}$ to obtain the generating lattice

\begin{align}
 \mathcal{L}:=\ker_{\mathbb{Z}}\sfA.
\end{align}

\noindent Finally, we set $[\gamma]_{u_-}=0$ whenever $w.u<0$ (see Lemma 3.17 in Ref.\cite{cattani2006}). Clearly, this property 
allows us to choose certain weights $w$ that simplify the sums.

We can use \textsl{Macaulay2} to perform the above operations. An example is provided in Appendix \ref{glossary}.

\subsection{Examples}
\label{examples-math}
In order to illustrate the methods, we will take two representations of the  Gauss hypergeometric function.

\subsubsection{Double integral Gauss hypergeometric function}

The first part of this example follows  \cite{2010arXiv1010.5060N}, where the reader can  find details. Suppose
we are interested in the following integral

\begin{align}
I(\alpha)= \int_{\mathbb{R}_+^2} \frac{z_1^{\alpha_1} z_2^{\alpha_2}}{(1+z_1+z_2+c z_1 z_2)^\beta} 
\frac{\dd z_1 \dd z_2}{z_1 z_2}.
\end{align}

\noindent From the point of view we are adopting, we should think on the more general situation where the 
polynomial in the denominator has indeterminate coefficients, i.e., 

\begin{align}
b(c, z)=c_1+ c_2 z_1+ c_3 z_2+c_4 z_1 z_2.
\end{align}

\noindent Its  associated configuration matrix reads

\begin{align}
\mathsf{A}= \begin{pmatrix}
    1&1&1&1\\
    0&1&0&1\\
    0&0&1&1
   \end{pmatrix}.
      \label{2F1-conf}
\end{align}

\noindent Then we consider the integral 

\begin{align}
I_b(-\beta, -\alpha)= \int_{\Omega} \frac{z_1^{\alpha_1} z_2^{\alpha_2}}{(c_1+ c_2 z_1+ c_3 z_2+c_4 z_1 z_2)^\beta} 
\frac{\dd z_1 \dd z_2}{z_1 z_2},
\end{align}

\noindent where $\Omega$ is a suitable cycle of integration. Following \cite{2010arXiv1010.5060N} the Newton 
polytope of $b(c, z)$ can be represented by the inequalities(See Eq.\eqref{polytope-half-space})

\begin{align}
 \Delta_b= \{\sigma_1\ge 0\}\cap \{\sigma_2\ge 0\}\cap \{-\sigma_1\ge -1\}\cap \{-\sigma_2\ge -1\},
\end{align}
\noindent and hence, we can read off the vectors $\mu_i$ and numbers $\nu_i$

\begin{align}
\mu_1&=(1,0),\quad \mu_2=(0, 1), \quad  \mu_3=(-1,0),\quad  \mu_4=(0,-1), \nonumber \\
\nu_1&= 0,\quad  \nu_2=0,\quad  \nu_3=-1, \quad  \nu_4=-1.
\end{align}

\noindent Integration cycles can be obtained taking $\theta=(\arg(c_1/c_2), \arg(c_1/c_3))$. The BFP
Theorem states that

\begin{align}
I_b(-\beta, -\alpha)=  \Phi^{\Theta}_b(\alpha, \beta,c) \Gamma(\alpha_1) \Gamma(\alpha_2)
\Gamma(\beta-\alpha_1) \Gamma(\beta-\alpha_2),
\label{integral-2f1}
\end{align}

\noindent where $\Theta\in \mathbb{T}^2\backslash \overline{\mathcal{A}'_b}$. Taking the representative $\theta\in
\Theta$, the entire function reads

\begin{align}
 \Phi^{\Theta}_b(\alpha,\beta, c)= \frac{c_1^{\alpha_1+\alpha_2-\beta} c_2^{-\alpha_2} 
 c_ 3^{-\alpha_3}}{\Gamma(\beta)^2}\; {}_2F_1(\alpha_1,\alpha_2,\beta;1-\frac{c_1 c_4}{c_2 c_3}).
\end{align}

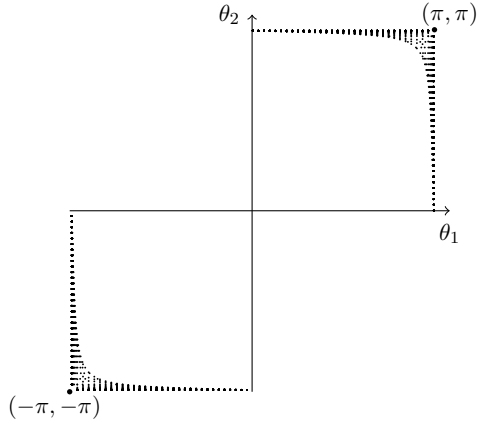
\begin{figure}[ht]
\centering
\begin{tikzpicture}
\node[anchor=south west,inner sep=0] at (0,0)
{\includegraphics[width=0.3\textwidth]{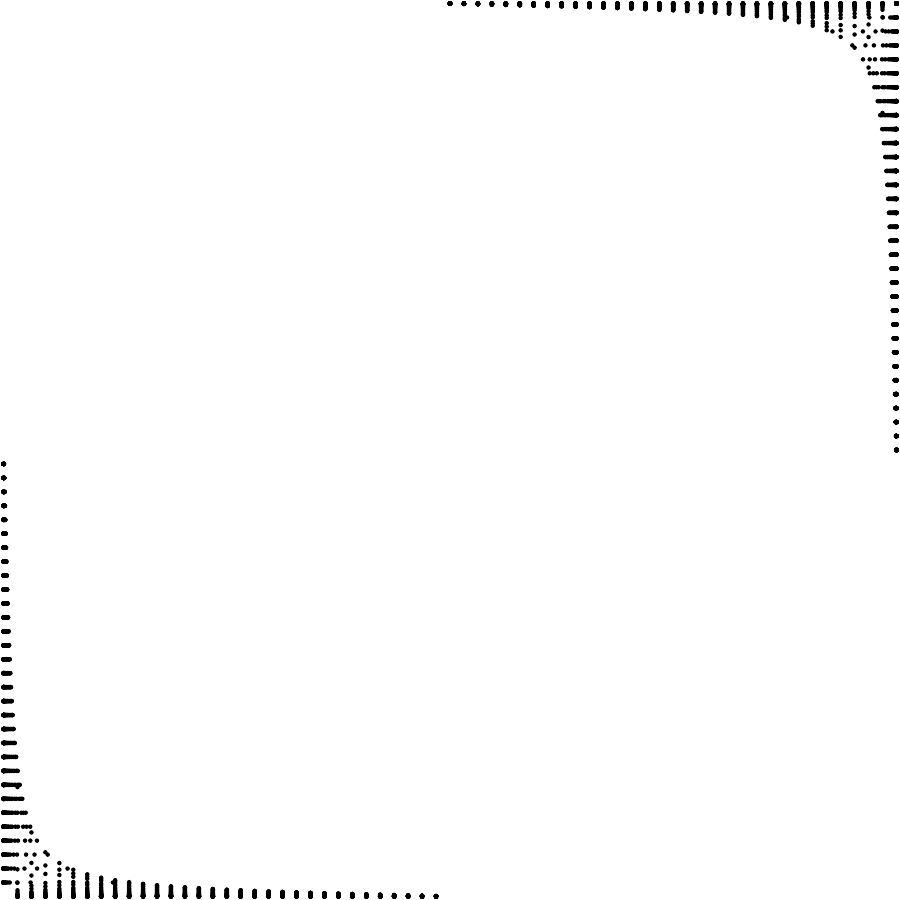}};
\draw [->](0,2.4) -- (5.0,2.4);
\draw [->](2.4,0) -- (2.4,5.0);
\node[scale=0.8, text centered] at (-0.2,-0.2) {$(-\pi, -\pi)$};
\node[scale=0.8, text centered] at (5,5) {$(\pi,\pi)$};
\node[scale=0.8, text centered] at (2.1, 5) {$\theta_2$};
\node[scale=0.8,text centered] at (5,2.1) {$\theta_1$}; 
\draw (4.8,4.8) node[fill,circle, inner sep=0pt, minimum size=2pt] {};
\draw (0,0) node[fill,circle, inner sep=0pt, minimum size=2pt]{};
\end{tikzpicture}
\caption{A connected component of the coamoeba(shaded) of $b(z)=1+z_1+z_2+c z_1 z_2$ drawn in the fundamental
domain $[-\pi, \pi] \times  [-\pi, \pi]$ of $\mathbb{T}^2$ in $\mathbb{R}^2$. The coefficient $c$ is near $1$.}
\label{coamoeba2f1}
\end{figure}

\noindent Taking the point $c'=(1,1,1,c)$, we have the coamoeba of  $b(z)$ shown in Fig.\ref{coamoeba2f1}, where
we see that  $(0,0)\notin \overline{\mathcal{A}'_b}$. There is one connected component of
$\mathbb{R}^2\backslash \overline{\mathcal{A}'_b}$  at this point,
therefore we have a single solution of $H_\sfA(-\beta,-\alpha)$. Notice that  at 
$c'$, we have $\Omega=\text{Arg}^{-1}\theta=\text{Arg}^{-1}(\arg(c_1/c_2), \arg(c_1/c_3))= \text{Arg}^{-1}(0, 0)=(0,\infty)^2$. 
Hence, at $c'$ we have

\begin{align}
 I(\alpha)=\frac{\Gamma(\alpha_1) \Gamma(\alpha_2)
\Gamma(\beta-\alpha_1) \Gamma(\beta-\alpha_2)}{\Gamma(\beta)^2}\; {}_2F_1(\alpha_1,\alpha_2,\beta;1-c).
\end{align}

It is instructive to recover this result from the canonical series algorithm. Let  $w=(0,1,1,1)$\footnote{Choices of $w$ 
are tied with the property that for $w\cdot u <0$ we have $[\gamma]_{u_-}=0$. See Algorithm in Sec.\ref{SST-algorithm}.} and 
 $\kappa=(-\beta, -\alpha_1,-\alpha_2)$. The toric ideal associated with $\sfA$ reads

\begin{align}
I_\sfA=\braket{\partial_2\partial_3-\partial_1\partial_4}. 
\end{align}

\noindent We have $\text{in}_w(I_\sfA)=\braket{\partial_2 \partial_3}$, which we use to obtain its  standard pairs

\begin{align}
 \mathcal{S}(\text{in}_w(I_\sfA))= \{(1, \{1,3,4\}), (1, \{1,2,4\})\},
\end{align}

\noindent therefore
\begin{align}
  \text{ind}_w(I_A)=& \braket{\theta_2}\cap \braket{\theta_3}.
\end{align}
\noindent Thus, we  obtain 
\begin{align}
 \finw=& \braket{\theta_2\theta_3, \theta_1+\theta_2+\theta_3+\theta_4+\beta,
 \theta_2+\theta_4+\alpha_1,\theta_3+\theta_4+\alpha_2}.
\end{align}

\noindent Computing its roots leads to 

\begin{align}
 \{\gamma_i\}=\{(\alpha _1-\beta ,\alpha _2-\alpha _1,0,-\alpha _2),
 (\alpha _2-\beta ,0,\alpha _1-\alpha _2,-\alpha _1)\}.
\end{align}

\noindent In addition, computing $\ker(\sfA)$ we have $\mathcal{L}=\mathbb{Z}(1,-1,-1,1)$. Let $u=n(1,-1,-1,1)$,
then $u.w=-n$, which means that $[\gamma]_{u_-}=0$ for $n>1$. In order to start the sum from $n=0$, we set
$u=n(-1,1,1,-1)$ such that  $[\gamma]_{u_-}=0$ for $n<0$. We then write 

\begin{align}
u= u_+-u_-=(0,n,n,0)-(n,0,0,n).  
\end{align}

\noindent Let us work out the case of $\gamma_1$. Eq.\eqref{log-free-series} gives  

\begin{align}
\phi_{1}=&\sum\limits_{n=0}^\infty \frac{[\gamma_1]_{u_-}}{[\gamma_1+u]_{u_+}} (c^{u+\gamma_1}) 
=c^{\gamma_1} \sum\limits_{n=0}^\infty \frac{[(\alpha _1-\beta ,\alpha _2-\alpha _1,0,-\alpha _2)]_{u_-}}{[(\alpha _1-\beta ,\alpha _2-\alpha _1,0,-\alpha _2)
+n(-1,1,1,-1)]_{u_+}} (c^{u}),  
\end{align}

\noindent hence

\begin{align}
\phi_{1}=&c^{\gamma_1}\sum\limits_{n=0}^\infty  \frac{(\beta-\alpha_1)_n(\alpha_2)_n}
{\poch{\alpha_2-\alpha_1+1} \poch{1}}\left(\frac{c_2 c_3}{c_1 c_4}\right)^n=c^{\gamma_1} {}_2F_1(\beta-\alpha_1,\alpha_2;
\alpha_2-\alpha_1+1;(c_2 c_3)/(c_1 c_4)),
\end{align}

\noindent where we have used the series representation of the Gauss hypergeometric function
\eqref{2f1-series-sum}. Similarly, for
$\gamma_2$ we have 

\begin{align}
\phi_{2}=&c^{\gamma_2}\sum\limits_{n=0}^\infty  \frac{(\beta-\alpha_2)_n(\alpha_1)_n}
{\poch{\alpha_1-\alpha_2+1} \poch{1}}\left(\frac{c_2 c_3}{c_1 c_4}\right)^n
=c^{\gamma_2} {}_2F_1(\beta-\alpha_2,\alpha_1;
\alpha_1-\alpha_2+1;(c_2 c_3)/(c_1 c_4)).
\end{align}

\noindent Therefore, the solution as linear combination of canonical series reads

\begin{align}
  I_b(-\beta, -\alpha)=& K_1   c^{\gamma_1}\ {}_2F_1(\beta-\alpha_1,\alpha_2;
\alpha_2-\alpha_1+1;(c_2 c_3)/(c_1 c_4))  \label{integral-2f1-canonical}\\
&+K_2 c^{\gamma_2} \ {}_2F_1(\beta-\alpha_2,\alpha_1;
\alpha_1-\alpha_2+1;(c_2 c_3)/(c_1 c_4)).\nonumber
\end{align}

\noindent We proceed to compute the integration constants. We take the noncompact cycle $\Omega=\mathbb{R}^2_+$. From
the roots, we observe that  if $c_3=0$, then $\phi_{2}$ vanishes and if
 $c_2=0$,  $\phi_{1}$ vanishes. Hence we make
 
\begin{align}
\int_{\mathbb{R}_+^2} \frac{z_1^{\alpha_1} z_2^{\alpha_2}}{(c_1+ c_2 z_1+c_4 z_1 z_2)} 
\frac{\dd z_1 \dd z_2}{z_1 z_2}=\left(K_1 \phi_{1}+K_2 \phi_{2}\right)|_{c_3\rightarrow0},
\end{align}
\noindent which leads to 
\begin{align}
 K_1= \frac{\Gamma(\alpha_2)\Gamma(\alpha_1-\alpha_2)\Gamma(\beta-\alpha_1)}{\Gamma(\beta)}.
\end{align}
\noindent Similarly, we obtain
\begin{align}
 K_2= \frac{\Gamma(\alpha_1)\Gamma(\alpha_2-\alpha_1)\Gamma(\beta-\alpha_2)}{\Gamma(\beta)}.
\end{align}

\noindent Let us now recover the original integrals by setting  $c_1=c_2=c_3=1$, $c_4=c$ 
in Eq.\eqref{integral-2f1-canonical}. We have

\begin{align}
I(\alpha)=& c^{-\alpha_2}
 \frac{\Gamma(\alpha_2)\Gamma(\alpha_1-\alpha_2)\Gamma(\beta-\alpha_1)}{\Gamma(\beta)} {}_2F_1(\beta-\alpha_1,\alpha_2;\alpha_2-\alpha_1+1;1/c)\\
&+c^{-\alpha_1}\frac{\Gamma(\alpha_1)\Gamma(\alpha_2-\alpha_1)\Gamma(\beta-\alpha_2)}{\Gamma(\beta)}
{}_2F_1(\beta-\alpha_2,\alpha_1;\alpha_1-\alpha_2+1;1/c), \nonumber
\end{align}

\noindent which equals Eq.\eqref{integral-2f1} after using the identity \eqref{id3} with $z=1-c$.

\subsubsection{Single integral Gauss hypergeometric function}

Let us now study the univariate integral of the Gauss hypergeometric function. We have 

\begin{align}
 I(-\beta_1,-\beta_2, -\alpha)= 
 \int\limits_{\mathbb{R}_+} \frac{z^\alpha}{(1+z)^{\beta_1} (1+ cz)^{\beta_2}} \frac{\dd z}{z}.
 \label{integral2F1-nontoric}
\end{align}

\noindent In order to evaluate the integral we consider the (toric) polynomial

\begin{align}
 b(z)^\beta= (c_1+c_2 z)^{\beta_1} (c_3+c_4 z)^{\beta_2} \Longleftrightarrow 
  \mathsf{A}=
\begin{pmatrix}
 1 & 1 & 0 & 0 \\
 0 & 0 & 1 & 1 \\
 0 & 1 & 0 & 1 \\
\end{pmatrix}
\end{align}

\noindent and its associated GKZ system. The integral under consideration reads

\begin{align}
 I_b(\kappa)= 
 \int\limits_{\Omega} \frac{z^\alpha}{(c_1+ c_2z)^{\beta_1} (c_3+ c_4z)^{\beta_2}} \frac{\dd z}{z},
  \label{integral2F1}
\end{align}
\noindent where $\kappa=(-\beta_1,-\beta_2, -\alpha)$. We have $\mathcal{L}=\mathbb{Z}(1,-1,-1,1)$. Taking the weight vector 
 $w=(0,1,1,1)$, we obtain 	

\begin{align}
\finw=&\braket{\theta _2 \theta _3,\beta _1+\theta _1+\theta _2,\beta _2+\theta _3+\theta _4,\alpha +\theta _2+\theta _4},\\
 \{\gamma_i\}=&\{(-\beta _1,0,\alpha-\beta_2,-\alpha),
 (\alpha -\beta_1-\beta_2,\beta_2-\alpha,0,-\beta_2)\}. 
\end{align}

\noindent Hence, we have the two series solutions

\begin{align}
 \phi_{1}=& c^{\gamma_1} {}_2F_1(\beta_1, \alpha; \alpha-\beta_2+1; 
 \frac{c_2 c_3}{c_1c_4}),\\
 \phi_{2}=& c^{\gamma_2} {}_2F_1(\beta_1+\beta_2-\alpha, \beta_2;
 \beta_2-\alpha+1; \frac{c_2 c_3}{c_1c_4}).
\end{align}

\noindent Specializing to our case, we have $c_1=c_2=c_3=1$ and $c_4=c$, therefore the 
solution of the integral is given by 

\begin{align}
I(-\beta_1,-\beta_2, -\alpha)&=K_1 \ c^{-\alpha}
{}_2F_1(\beta_1, \alpha; \alpha-\beta_2+1;1/c)\\
&+K_2 \ c^{-\beta_2}\ {}_2F_1(\beta_1+\beta_2-\alpha, \beta_2;\beta_2-\alpha+1;1/ c),\nonumber
\end{align}

\noindent where the constants can be obtained by setting  $c_2$ and $c_3$ to zero in Eq.\eqref{integral2F1} choosing
$\Omega=\mathbb{R}_+$. The final result reads

\begin{align}
 I(-\beta_1,-\beta_2, -\alpha)=& \frac{\Gamma(\alpha)\Gamma(\beta_2-\alpha)}{ \Gamma(\beta_2)} c^{-\alpha}
{}_2F_1(\beta_1, \alpha; \alpha-\beta_2+1;1/c) \\
&+ c^{-\beta_2} \frac{\Gamma(\alpha-\beta_2)\Gamma(\beta_1+\beta_2-\alpha)}{\Gamma(\beta_1)}
 {}_2F_1(\beta_1+\beta_2-\alpha, \beta_2;\beta_2-\alpha+1; 1/c),\nonumber
\end{align}

\noindent which evaluates to \eqref{integral2F1-nontoric} after using Eq.\eqref{id3} with $z=1-c$.

\section{Feynman integrals as $\sfA$-hypergeometric functions}
\label{sec3}

In this Section we will interpret Feynman integrals as particular points of $\sfA$-hypergeometric functions. First, we
will introduce the parametric representation based on $g=\mathcal{U}+\mathcal{F}$ used in Ref.\cite{Lee:2013hzt} by Lee 
and Pomeransky. Later, we will give our proposal for defining GKZ systems based on $g$. Examples will be presented at the  
end of this Section.

\subsection{Lee-Pomeransky representation of Feynman integrals}
Let us consider a typical  Feynman  integral in Euclidean space in dimensional regularization. 
A $L$-loop integral with $N$ propagators and $E$ independent external momenta may be written
as follows

\begin{align}
 I_F(\alpha)= \int\limits_{\mathbb{R}^L} \left(\prod\limits_{i=1}^L \frac{\dd^d k_i}{\pi^{d/2}}\right)
 \frac{1}{D_{1}^{\alpha_1} \cdots D_{N}^{\alpha_N}}, \label{FeynmanIntegral} 
\end{align}

\noindent where  the inverse propagators are of the form 

\begin{align}
 D_i= (M_i)^{rs}k_r\cdot k_s + 2 (Q_i)^{rs} k_r\cdot p_s+J_i.  
\end{align}

\noindent The matrices, $M_i$, $Q_i$, and $J_i$ have dimensions $L\times L$, $L\times E$, and 
$1\times 1$, respectively. Integration over loop momenta through Feynman parameters and a Mellin
transform leads to the Lee-Pomeransky parametric representation \cite{Lee:2013hzt} 

\begin{align}
 I_F(\alpha)= \xi_{\Gamma_{\alpha}} 
 \int_{\mathbb{R}_+^N}\left( \prod\limits_{i=1}^N \frac{\dd z_i}{z_i} z_i^{\alpha_i}\right) \frac{1}{g(z)^{d/2}}
  \label{Leerep}= \xi_{\Gamma_{\alpha}} 
 \int_{\mathbb{R}_+^N} \frac{z^\alpha}{g(z)^{d/2}}  \dd \eta_N,
\end{align} 
\noindent  where we have used the multi-index notation  in the second equality and $\mathbb{R}_+=(0,\infty)$. The overall factor and the
polynomial $g(z)$ are defined as follows:

\begin{align}
 \xi_{\Gamma_{\alpha}}:=& \frac{\Gamma(d/2)}{\Gamma((L+1)d/2-\sum_{i=1}^N \alpha_i) \prod_{i=1}^N \Gamma(\alpha_i)},\\
 g(z):=&\mathcal{U}+\mathcal{F},
\end{align}

\noindent where $\mathcal{U}$ and $\mathcal{F}$ are the Symanzik polynomials. In order to compute them, we construct the matrices

\begin{align}
 M^{rs}= \sum\limits_{i=1}^N z_i M_i^{rs}, \quad  Q^{r}= \sum\limits_{i=1}^N z_i Q_i^{rs}p_s,\quad   
  J=\sum\limits_{i=1}^N z_i J_i.
\end{align}

\noindent We then have 

\begin{align}
 \mathcal{U}= \det(M), \quad \mathcal{F}=\det(M)\left(J-\left(M^{-1}\right)^{ij} Q^{i}\cdot Q^{j} \right), 
\end{align}

\noindent where  $\mathcal{F}$ is appropriately scaled in order to make it dimensionless.  
The polynomial $\mathcal{U}$ is a homogeneous polynomial of degree $L$ and the polynomial $\mathcal{F}$
is homogeneous of degree $L+1$. Hence $g(z)$ is an inhomogeneous polynomial of degree $L+1$. In Euclidean
kinematics, the Symanzik  polynomials $\mathcal{U}$, $\mathcal{F}$ are positive semi-definite functions of the Feynman parameters.
These polynomials can also be obtained from the topology of the graphs and their properties are 
summarized in Refs.\cite{Bogner:2007cr, Bogner:2010kv, Weinzierl:2013yn}.

\subsection{Feynman integrals and canonical series}

A suitable definition for computational algebra purposes would be to consider Feynman integrals as a linear 
combinations of their canonical series, whenever we can compute them. As they stand, Feynman integrals
will not satisfy the system of PDEs associated with a GKZ system (Eqs.\eqref{PDE1}-\eqref{PDE2})
because the polynomial $g(z)$ has fixed coefficients.  

Accordingly, the first step in our construction will be to take $g(z)$ and consider its associated toric polynomial 
$g(c,z)$. We will demand that this polynomial furnish GKZ system in the sense of  Eq.\eqref{log-free-series}. In particular, this means that $g(c,z)$ must lead to matrices of 
$\text{co}(\sfA)>0$ in order to compute a generating lattice and follow the canonical series algorithm.  Since we 
have a single polynomial, the codimension of the matrix $\sfA$ associated with $g(c,z)$ is given by 
$\text{co}(\sfA)=n-N-1$, where $n$ is the number of monomial terms in $g(c, z)$ with common exponent vectors. Therefore,
we will have $\text{co}(\sfA)=0$ when the number of terms in $g(c, z)$ equals the number of rows in $\sfA$, or equivalently when $n=N+1$. A
typical example of this situation is the polynomial of the  $L$-loop massless cantaloupe graph(Fig.\ref{cantaloupe-diagram}). 
For instance, let us consider the case $L=1$, i.e., the massless bubble graph (Fig.\ref{massless-bubble}). 
The $g(z)$ polynomial of this graph is given by $g(z)=z_1+z_2+s z_1 z_2$, therefore  

\begin{align}
 g^{\text{bubble}}(c, z)=c_2 z_1 +c_3 z_2 + c_4 z_1z_2 \Longleftrightarrow \sfA^{\text{bubble}}=\begin{pmatrix}
                                                              1&1&1\\
                                                              1&0&1\\
                                                              0&1&1
                                                             \end{pmatrix}.
                                                             \end{align}

\noindent The codimension of $\sfA^{\text{bubble}}$ is zero and hence $\text{ker } \sfA= \emptyset$. Therefore,
we cannot  use this matrix to represent a solution of a GKZ system as canonical series in the sense of Eq.\eqref{log-free-series}. However, in this simple 
 example we know that  the result of the massless bubble integral can be recovered from the one-mass bubble integral by taking the limit of
the mass going to zero. Furthermore, as we will see in the examples, the matrix associated with the one-mass 
bubble integral  is given by 

\begin{align}
\sfA^{\text{one-mass}}= 
\left(
\begin{array}{cccc}
 1 & 1 & 1 & 1 \\
 1 & 0 & 1 & 0 \\
 0 & 1 & 1 & 2 \\
\end{array}
\right),
\end{align}

\noindent which has $\text{co}(\sfA^{\text{one-mass}})=1$.  Since the one-mass bubble evaluates to a Gauss hypergeometric function,
we may consider the massless bubble as a special case of a ${}_2F_1(a,b,c;p^2/m^2)$ function, for some
$a,b,c$ depending on the powers of the propagators and the dimension. An equivalent alternative provided by GKZ systems is to deform 
the polynomial by adding  an arbitrary constant $r(z)=c_1$ to $g^{\text{bubble}}(c, z)$ and 
consider instead

\begin{align}
 g_r(c, z)=c_1+c_2 z_1 +c_4 z_2 + c_4 z_1z_2 \Longleftrightarrow \sfA^r=\begin{pmatrix}
                                                              1&1&1&1\\
                                                              0&1&0&1\\
                                                              0&0&1&1
                                                             \end{pmatrix},
                                                             \end{align}

\noindent where the constant $c_1$ can be set to zero at the end of the computation. This matrix corresponds to the 
GKZ system of a Gauss hypergeometric function. We will work  out explicitly this example in 
Section \ref{massless-bubble-ex}. There, we will take  the limit $c_1\rightarrow 0$ of a Gauss hypergeometric 
function as we would do for the mass in the case of a one-mass bubble. The choice of a deformation can be made 
systematically as we will see for the  $L$-loop cantaloupe graph.

Therefore, in cases where $g(c, z)$ leads to a configuration matrix of $\text{co}(\sfA)=0$, we will consider a 
deformation of $g(c, z)$ demanding that it defines a solution of a  GKZ system in the sense of Eq.\eqref{log-free-series}, 
thus allowing us to construct canonical series solutions. Recall that $\text{deg }(\mathcal{U})=L$ and 
$\text{deg }(\mathcal{F})=L+1$. We may choose any polynomial $r(z)$ of $\text{deg} (r)<L$ and 
define

\begin{align}
g_r(c, z):=r(c, z)+\mathcal{U}(c)+\mathcal{F}(c), 
\end{align}

\noindent with the requirement that the associated matrix has $\text{co}(\sfA)>0$.
Here $r(c,z)$ denotes the toric polynomial associated with $r(z)$. Similarly, $\mathcal{U}(c)$ and $\mathcal{F}(c)$ 
denote the Symanzik polynomials where the coefficients now are considered as variables.  We have checked for the 
integral of a massless $L$-loop cantaloupe graph, up to $5$-loop, 
that this definition leads to  well behaved GKZ systems and that we can recover the original integral using canonical 
series. 

The second step in our construction will be to consider Euler-type solutions and series solutions
of the GKZ systems furnished with $g_r(c,z)$. Let us first define the class of Euler integrals associated with the Feynman
integral \eqref{Leerep}. We will drop the overall factor $\xi_{\Gamma_\alpha}$, which  is  a nonzero 
constant independent of the kinematics and instead consider 

\begin{align}
I_F(\alpha)/\xi_{\Gamma_{\alpha}}\longmapsto I_{g_r}(-d/2,-\alpha):=
\int_{\Omega} \frac{ z^\alpha }{g_r(c, z)^{d/2}} \dd \eta_N,
\end{align}

\noindent which is a toric generalization of Eq.\eqref{Leerep}, i.e., where the deformation has been 
introduced and the coefficients of $g_r(z)$ are considered as variables. Notice that at this point, 
the noncompact cycle $\Omega$ will not be in general $\mathbb{R}_+^N$ and it will be determined 
by the coamoeba of $g_r(c,z)$ (Sec.\ref{Euler-type-integrals}).

Let us first show that the class of integrals obtained from $g_r(c, z)$ is $\sfA$-hypergeometric\footnote{A related 
construction---which does not introduce deformations---has been considered in Ref.\cite{2016arXiv160504970N} for the 
special case where the  powers of the propagators are given by  $\alpha=(1,\dots, 1)$. Here we consider generic powers in the propagators as dictated by canonical series solutions.}. 
We have the following theorem, which is a specialization of theorem 2.7 in Ref.\cite{GELFAND1990255} for the
case of a single polynomial $g_r(c, z)$. That this theorem still holds for noncompact cycles is 
discussed in Ref.\cite{2011arXiv1103.6273B,2017arXiv170303036F}.   

\subsubsection*{Theorem.}
\emph{Let $g_r(c, z)$ be the deformed polynomial in $N$ variables obtained from 
$g(c, z)=\mathcal{U}(c)+\mathcal{F}(c)$, where $\mathcal{F}(c)$ and  $\mathcal{U}(c)$ are obtained by considering
the coefficients appearing in the Symanzik polynomials as variables. $g_r(c, z)$ is obtained by introducing a 
deformation $r(c,z)$  demanding that its matrix satisfies $\text{co}(\sfA)>0$.  Let 
$A=(a_1 \ a_2 \cdots  \ a_n)$ be the configuration matrix
associated with $g_r(c, z)$ and consider the polynomial with indeterminate generic coefficients}

\begin{align}
g_r(c, z)  = \sum_{i=1}^{n} c_{i} z^{a_i}, \qquad c_{i}\in\mathbb{C}_{*}.
\end{align}

\noindent \emph{ Let $\mathsf{A}$ be its associated $(N+1) \times n$ matrix}

\begin{align}
\mathsf{A}=
\begin{pmatrix}
1&1&\dots& 1\\
a_1 & a_2 & \dots & a_n
\end{pmatrix}.\label{config-gr}
\end{align}

\noindent \emph{The Euler-Mellin integral }
\begin{align}
 I_{g_r}(\kappa) =  \int_{\Omega} \frac{ z^\alpha }{g_r(c, z)^{d/2}} \dd \eta_N \label{fintegral-A-hypergeometric}
\end{align}

\noindent \emph{is a solution of the $\sfA$-hypergeometric  system $H_{\mathsf{A}}(\kappa)$ of degree 
$\kappa=(-d/2, -\alpha)$. Noncompact cycles $\Omega$ can be obtained by taking the coamoeba 
of $g_r(c, z)$ and choosing representatives $\theta$ of connected components $\Theta \in \mathbb{R}^N\backslash 
\overline{\mathcal{A}'}_{g_r}$(See Sec.\eqref{Euler-type-integrals}.)}

\noindent\textbf{Proof.} The proof goes along the lines of
Ref.\cite{2017arXiv170303036F} specializing  to the nonhomogeneous case and 
the case of a single polynomial.  Let us consider  first Eq.\eqref{PDE1}. We have

\begin{align}
 (\partial_1)^{u_1} (\partial_2)^{u_2} \cdots (\partial_n)^{u_n}  I_{g_r}(\kappa)&=  (-d/2)
  (\partial_1)^{u_1} (\partial_2)^{u_2} \cdots  (\partial_{n})^{u_{n}-1}
 \int_{\Omega} \dd \eta_N z^\alpha g_r(c , z)^{-d/2-1} z^{a_i} \nonumber\\
 &= (-d/2)_{(|u_n|)} (\partial_1)^{u_1} (\partial_2)^{u_2} \cdots  (\partial_{n-1})^{u_{n-1}}\!\!\!
 \int_{\Omega}\dd \eta_N z^\alpha  g_r(c, z)^{-d/2-|u_n|} z^{|u_n| a_n} \nonumber\\
&= (-d/2)_{(|u_1|+\dots +|u_n|)} \int_{\Omega} \dd \eta_N z^\alpha  
g_r(c, z)^{-d/2-|u_1|-\dots -|u_n|} 
z^{|u_1| a_1 + \dots +|u_n| a_n}\nonumber,
\end{align}

\noindent where $(\rho)_{(x)}$ denotes the falling factorial. Similarly  

\begin{align}
 (\partial_1)^{v_1} (\partial_2)^{v_2} \cdots (\partial_n)^{v_n}  I_{g_r}(\kappa)=
 & (-d/2)_{(|v_1|+\dots +|v_n|)} \int_{\Omega} \dd \eta_N z^\alpha  
g_r(c, z)^{-d/2-|v_1|-\dots -|v_n|} 
z^{|v_1| a_1 + \dots +|v_n| a_n},\nonumber
\end{align}

\noindent and from $\sfA u=\sfA v$, the result $(\partial^u-\partial^v)I_{g_r}(\kappa)=0$ follows.  

Let us now focus on the second set of differential equations \eqref{PDE2}. Consider the first row of $\sfA$ in
Eq.\eqref{config-gr}. We have 

\begin{align}
( c_1 \partial_1 +  c_2 \partial_2 +
 \dots +c_n \partial_n )I_{g_r}(\kappa)=& 
 \int_{\Omega} \dd \eta_N z^\alpha (-d/2) g_r(c, z)^{-d/2-1} (c_1 z^{a_1}+\dots + 
 c_n z^{a_n} )\nonumber\\
 =&(-d/2) I_{g_r}(\kappa).\nonumber
\end{align}

\noindent Similarly, for $i>1$

\begin{align}
\sum\limits_{j=1}^n a_{ij} c_j\partial_j  I_{g_r}(\kappa)
 =&\sum\limits_{j=1}^n a_{ij}  \int_{\Omega} \dd \eta_N z^\alpha (-d/2)
g_r(c, z)^{-d/2-1} (c_j z^{a_j})\nonumber\\
=& \int_{\Omega} \dd \eta_N z^\alpha (-d/2)
g_r(c,z)^{-d/2-1} \left( z_{i} \frac{\partial}{\partial z_{i}} g_r(c, z)\right) \nonumber\\
= & \int_{\Omega} \dd \eta_N z^\alpha \left(z_{i}
\frac{\partial}{z_{i}} g_r(c, z)^{-d/2}\right)\nonumber\\
=& - \alpha_i I_{g_r}(\kappa), \nonumber
\end{align}

\noindent where we have used integration by parts in the last equality. This completes 
the proof.

In this way, we have generated a class of integrals related with the Lee-Pomeransky representation of 
 Feynman integrals. Feynman integrals will correspond to special cases (points) of  $\sfA$-hypergeometric functions 
 whenever the cycle $\Omega$ can be taken as $\mathbb{R}_+^N$ and the coefficients $c$ can be taken as functions of 
 the kinematic invariants.  The behavior of the integral as the coefficients $c$ vary can be studied through Eq.\eqref{integral-coamoeba}. This representation is also 
 useful as it provides noncompact cycles at the cost of imposing conditions on $\kappa$. However, this can
 be sorted out by analytic continuation and it can be shown that, as a function of the  \emph{variables} $c$,
 Eq.\eqref{integral-coamoeba} is $\sfA$-hypergeometric everywhere. Here $c$ is taken in
 $\mathbb{C}^n\backslash \Sigma_\sfA$, where $\Sigma_\sfA$
 denotes the singular locus of all $\sfA$-hypergeometric functions(Theorem 4.2 in \cite{2011arXiv1103.6273B}). The main
 difficulty in this  approach is to choose a representative of some connected component $\Theta$ in 
 $\mathbb{R}^N\backslash\overline{\mathcal{A}'_{g_r}}$. In other 
 words, we have to select a point such that the cycle is nonvanishing on the set $\text{Arg}^{-1}(\theta)$,
 where $\theta\in \Theta$.  The integration region of the Feynman integral, namely $\mathbb{R}^N_+$, suggest  taking any
 $\theta=( \arg(f_1(c)), \dots,\arg(f_N(c)))$, $f_i(c)>0$ provided $\theta \notin \overline{\mathcal{A}'_{g_r}}$
 (See e.g. Fig.\ref{coamoeba2f1}). This gives one connected component of $\text{vol}(\sfA)$ many ones 
 and a possible integration cycle\footnote{For all real positive coefficients in $g(z)$, i.e., the nontoric polynomial, 
 we can ensure that $0$ is in $\mathbb{T}^N$ \cite{2011arXiv1103.6273B}.}.   
 
 On the other hand, logarithm-free canonical series allows us to study  the behavior of the solution space 
 under variations of the parameters $\kappa$ at nonsingular points \cite{sturmfels:2001,2016arXiv160308954B}. 
 More important, once we have identified the above integrals as solutions of a GKZ system, we
 can use the statement in Eq.\eqref{linear-combination} relating Euler-type  integrals 
 and canonical series. For our case, this statement reads 
 
 \begin{align}
 \int_{\Omega} \frac{ z^\alpha }{g_r(c, z)^{d/2}} \dd \eta_N = K_1\phi_{1} +\dots +K_{M} \phi_{M},
  \label{CS-integral-constants}
 \end{align}

 \noindent where $\phi_1,\dots,\phi_M$  are  the canonical series associated with $\sfA$.  Fake exponents 
 $\gamma$ for each $\phi_1,\dots,\phi_M$ can be obtained from the SST algorithm in Sec.\ref{SST-algorithm}. 
 This equality is  fundamental  for our purposes as it allows both to take the limit of the deformation to zero and computing
 the integrations constants. Let us discuss these limits.
 
The form of the series solution for some fake exponent $\gamma$ reads

\begin{align}
 \phi_\gamma= c^\gamma \sum\limits_{u \in \mathcal{L}} \frac{[\gamma]_{u_-}}{[\gamma+u]_{u_+}} c^{u}. 
 \label{canonical-series-limit}
\end{align}

 \noindent These are characterized by a weight vector $w\in\mathbb{R}^n$, which selects a common domain
 of convergence $\mathcal{U}_w$ (See Theorem 2.5.16 in \cite{sturmfels:1999}). In addition, we have the restriction
 $[\gamma]_{u_-}=0$ for $u.w<0$. Taking the limit of the deformation to zero  amounts to take some of the 
 coefficients in $c=(c_1, \dots, c_n)$ to zero in Eq.\eqref{canonical-series-limit} and similarly for the  remaining  canonical series. On
 the LHS of Eq.\eqref{CS-integral-constants}, taking this limit  amounts to recover the undeformed integral 
 we started with. Let us clarify this point. The integral in the LHS of \eqref{CS-integral-constants} is a well defined
 $\sfA$-hypergeometric function provided $c$ are generic and the cycle is taken in some $\theta \notin \overline{\mathcal{A}'_{g_r}}$. The 
 RHS of \eqref{CS-integral-constants} is an asymptotic expansion of such integral. Therefore, the limits on the LHS
 correspond to special values of the $\sfA$-hypergeometric functions in the RHS. A judicious 
 choice of the deformation ensures that this limit can be taken systematically as we will see in the examples.
 
 In general,  this limit will not be smooth as it will require the evaluation of $\sfA$-hypergeometric functions on 
 their singular points\footnote{See Ref.\cite{2010arXiv1010.5060N} and the example in Sec.\ref{massless-bubble-ex}.}
 and therefore we may require analytic continuation of the corresponding $\sfA$-hypergeometric function. This can
 be seen as follows. A generic canonical series solutions can be understood as a function in $\text{co}(\sfA)$ variables.
 Setting one of these variables to zero may require a transformation of a variable to, say, its inverse and hence we require 
 analytic continuation. A judicious choice of the weight vector $w$ can simplify taking this limit, since it
 provides the condition $[\gamma]_{u_-}=0$ for $u.w<0$, thus determining the arrangement of the powers of $c^u$ in
 Eq.\eqref{canonical-series-limit}. As we will see for the $L$-loop cantaloupe graph, we can systematically 
 choose a deformation and a weight vector to take this limit.  However, we will leave as a conjecture that this this limit
 can always be taken.

 For integration constants, we will use the information available on the fake exponents following Ref.\cite{sturmfels:1999}.
 The initial series in Eq.\eqref{canonical-series-limit} are given by $c^\gamma$. Therefore, the positions of the $0$'s on 
 each $\gamma$ will tell us which elements in $c$ we have to set to zero in order to compute the integration constants.

Let us give some final comments. Notice that in order to recover the original Feynman integral---in agreement with our 
definition of the polynomials associated with the GKZ system and of the 
integral \eqref{fintegral-A-hypergeometric}---we must set coefficients $c$ to their kinematic values at the end of 
the computation. The integral \eqref{fintegral-A-hypergeometric} can be understood as a holomorphic function on
$\mathcal{U}_w$. This observation gives us a nice way of deducing  the long known fact that  convergent Feynman 
integrals are  functions of a Nilsson class \cite{Golubeva1976RuMaS}, which is clear  from their 
canonical series.

We end with our prescription to compute integration constants. 

\subsection*{Integration constants}

Suppose that we obtain $M$ fake exponents from the SST algorithm. Let us isolate an exponent, say, 
$\gamma=(\gamma^1, \dots, \gamma^n)$ and suppose it contains $k<n$ zeros in  positions  $\sigma_1, \dots, \sigma_k$. Take the 
coefficients associated with those positions to zero and consider

\begin{align}
  I_{g_r}(\kappa)\Bigg|_{\substack{c_{\sigma_1}
\rightarrow 0\\\vdots\\c_{\sigma_k}\rightarrow 0}}=
(K_1\phi_{1} +\dots +K_{M} \phi_{M})\Bigg|_{\substack{c_{\sigma_1}
\rightarrow 0\\\vdots\\c_{\sigma_k}\rightarrow 0}},  
\end{align}

\noindent where we take $\Omega=\mathbb{R}_+^{N}$ as the cycle in the LHS. This procedure will compute a single
coefficient. We repeat the process until we have computed all of them.

\subsection{Examples of co$(\sfA)=0$ }

The purpose of $\text{co}(\sfA)=0$ examples is to show how to deal with the appearance of a deformation.

\subsection*{Structure of the examples}

 In the following examples we will omit the overall gamma factors $\xi_{\Gamma_\alpha}$ and hence 
 consider
 
 \begin{align}
  I(\alpha):=I_F(\alpha)/\xi_{\Gamma_\alpha} \nonumber
 \end{align}

\noindent along with its toric  version $I_{g_{r}}(\kappa)$.  The vector $\kappa$ has the form 

\begin{align}
 \kappa=(-d/2, -\alpha_1, \dots, -\alpha_N), \nonumber
\end{align}

\noindent where $N$ is the number of propagators.  We assume that the powers of the propagators have generic 
noninteger complex values. This simplifies the discussion since the sum runs over
$\mathcal{L}=\ker_{\mathbb{Z}}\sfA$. The kernel of $\sfA$ leads to a rank $\text{co}(\sfA)$ lattice. For a single generator, we write 

\begin{align}
\mathcal{L}:=\mathbb{Z}(a_1, \dots, a_n),\quad  u=n(a_1, \dots, a_n),\quad  a_i \in \mathbb{Z}.  \nonumber
\end{align}
\noindent We set 
\begin{align}
\beta=d/2. \nonumber
\end{align}

\noindent In order to indicate the $i$-th component of a root vector $\gamma_r$ we
write $\gamma_r^i$.

\subsubsection{Massless bubble}
\label{massless-bubble-ex} 

The simplest example is the  massless bubble (Fig.\ref{massless-bubble}) with inverse propagators 
\begin{align}
 D_1=(k)^2, \qquad D_2=(k-p)^2,
\end{align}
\noindent where $s=-p^2$. We have 
\begin{align}
 g(z)=z_1+z_2+s z_1 z_2.
\end{align}
\noindent This polynomial leads to a matrix of codimension $\text{co}(\sfA)=0$, hence we introduce
a deformation $r(z)=c_1$ and consider instead 
\begin{align}
 g_r(c, z)= c_1+c_2z_1+c_3z_2+c_4 z_1 z_2 \Longleftrightarrow \sfA= \left(
\begin{array}{cccc}
 1 & 1 & 1 & 1 \\
 0 & 1 & 0 & 1 \\
  0 & 0 & 1 & 1\\
\end{array}
\right).
\end{align}
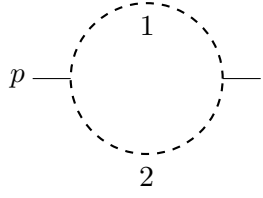
\begin{figure}[ht]
\centering
\begin{tikzpicture}
\draw [dashed, thick](0,1)circle (1);
\draw(-1.5,1)--(-1.0,1);
\draw(1.0,1)--(1.5,1);
\node[text width=0.5cm, text centered ] at (0,1.7) {$1$};
\node[text width=0.5cm, text centered ] at (0,-0.3) {$2$};
\node[text width=0.5cm, text centered ] at (-1.7,1.0) {$p$};
\end{tikzpicture} 
\caption{Bubble graph: $s=-p^2$}
\label{massless-bubble}
\end{figure}

\noindent Thus, we have

\begin{align}
 I_{g_r}(\kappa)=
 \int_{\text{Arg}^{-1}\theta} \frac{ z_1^{\alpha_1} z_2^{\alpha_2} }{(c_1+c_2z_1+c_3z_2+c_4 z_1 z_2)^{\beta}} 
 \frac{\dd z_1}{z_1} \frac{\dd z_2}{z_2},
\end{align}

\noindent for $\theta=(\arg(c_1/c_3),\arg(c_1/c_2))$. We have studied this integral in  Sec.\ref{examples-math}, 
where we established that

\begin{align}
I_{g_r}(\kappa) =\frac{ \Gamma(\alpha_1) \Gamma(\alpha_2)\Gamma(\beta-\alpha_1)  \Gamma(\beta-\alpha_2)}{\Gamma(\beta)^2}
c_1^{\alpha_1+\alpha_2-\beta} c_2 ^{-\alpha_1} c_3^{-\alpha_2} 
{}_2F_1(\alpha_1,\alpha_2;\beta;1-\frac{c_1 c_4}{c_2 c_3}).
\label{result-2fA-coamoeba}
\end{align}

\noindent In order to recover our Feynman integral we have to take the limit $c_1\rightarrow 0$.  The limit has to be 
taken carefully as $c_1$ appears both  as a factor and in the argument of the  hypergeometric function. Using the identity \eqref{id1} we have

\begin{align}
 {}_2F_1(\alpha_1,\alpha_2;\beta;1-\frac{c_1 c_4}{c_2 c_3})= \left(\frac{c_1c_4}{c_2 c_3}\right)^
 {(\beta-\alpha_1-\alpha_2)} {}_2F_1(\beta-\alpha_1, \beta-\alpha_2;\beta;
 1-\frac{c_1 c_4}{c_2 c_3}).
\end{align}

\noindent Therefore

\begin{align}
I_{g_r}(\kappa) =\frac{ \Gamma(\alpha_1) \Gamma(\alpha_2)\Gamma(\beta-\alpha_1)  \Gamma(\beta-\alpha_2)}
{\Gamma(\beta)^2}  c_2^{\alpha_2-\beta} c_3^{\alpha_1-\beta} c_4^{\beta-\alpha_1-\alpha_2} 
{}_2F_1(\beta-\alpha_1, \beta-\alpha_2;\beta;
 1-\frac{c_1 c_4}{c_2 c_3}). 
\end{align}

\noindent Finally, taking the limit $c_1=0$, setting $c_2=c_3=1$, and $c_4=s$ we recover the desired
result 

\begin{align}
 I(\alpha)= \frac{\Gamma(\beta-\alpha_1) \Gamma(\beta-\alpha_2)\Gamma(\alpha_1+\alpha_2-\beta)}{
 \Gamma(\beta)}s^{(\beta-\alpha_1-\alpha_2)},
\end{align}
\noindent where we have used   identity \eqref{id2}. Notice that in order to take this limit, we have evaluated
Eq.\eqref{result-2fA-coamoeba} on one of the singular points of the Gauss hypergeometric function, thus requiring 
an Euler transformation.

\subsubsection{The single-scale massless triangle graph}

\begin{figure}[ht]
\centering
\begin{tikzpicture}
%\draw[step=1cm,black,very thin] (-1.9,-1.9) grid (1.9,1.9);
\draw [dashed](0,-1) -- (1, 1);
\draw [dashed](0,-1) -- (-1, 1);
\draw [dashed](-1,1) -- (1, 1);
\draw[dashed](-1,1)--(-1.5,1.5);
\draw[dashed](1,1)--(1.5,1.5);
\draw(0,-1)--(-0.0,-1.5);
\node[text width=0.5cm, text centered ] at (-1,0) {$1$};
\node[text width=0.5cm, text centered ] at (1,0) {$2$};
\node[text width=0.5cm, text centered ] at (0,0.7) {$3$};
\end{tikzpicture}
\caption{Triangle graph: $s=-(p_1+p_2)^2$}
\label{triangle-graph}
\end{figure}
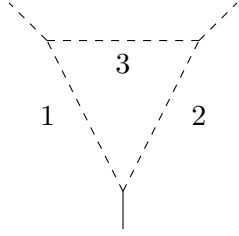

Let us now turn our attention to those limits through canonical series.  Let us consider the triangle graph in 
Fig.\ref{triangle-graph}. The inverse propagators read  

\begin{align}
D_1=(k_1 - p_1)^2,\qquad D_2=(k_1+p_2),\qquad D_3=k_1^2, 
\end{align}

\noindent where $-p_1^2=-p_2^2=0$.  Computing the relevant polynomial and taking the deformation 
$r(z)=c_1$ leads to  

\begin{align}
 g_r(c, z)= c_1+ c_2 z_1+ c_3 z_2+ c_4 z_3 + c_5 z_1 z_2 \Longleftrightarrow \sfA=\left(
\begin{array}{ccccc}
 1 & 1 & 1 & 1 & 1 \\
 0 & 1 & 0 & 0 & 1 \\
 0 & 0 & 1 & 0 & 1 \\
 0 & 0 & 0 & 1 & 0 \\
\end{array}
\right),
\end{align}

\noindent where, at the end of the computation, we will make the identifications $c_2=c_3=c_4=1$ and $c_5=s=-(p_1+p_2)^2$. 
The  integral reads
\begin{align}
 I_{g_r}(\kappa)=
 \int_{\Omega} \dd\eta_3 \frac{z_1^{\alpha_1} z_2^{\alpha_2} z_3^{\alpha_3}}{(c_1+ c_2 z_1+ c_3 z_2+ c_4 z_3 + c_5 z_1 z_2)^\beta}.
 \label{toric-triangle}
\end{align}

\noindent Computing $\ker (\sfA)$, we have 

\begin{align}
 \mathcal{L}=\mathbb{Z}(1,-1,-1,0,1) \Rightarrow u=n(1,-1,-1,0,1).
\end{align}

\noindent Choosing $w=(1,0,0,0,0)$, we have $u.w=n$, hence 
$[\gamma_i]_{u_-}=0 $ for $ n<0$. Setting $A=\alpha_1+\alpha_2+\alpha_3$, the fake indicial ideal 
and its roots read

\begin{align}
\finw=& \braket{\theta _1 \theta _5,\beta +\theta _1+\theta _2+\theta _3+\theta _4+\theta _5,\alpha _1+\theta _2+\theta _5,\alpha _2+\theta _3+\theta _5,
\alpha _3+\theta _4},\\
\{\gamma_i\}=& \{(0,\alpha _2+\alpha _3-\beta ,\alpha _1+\alpha _3-\beta ,-\alpha _3,-\alpha+\beta
),(A -\beta ,-\alpha _1,-\alpha _2,-\alpha _3,0)\}.                         
\end{align}

\noindent Inserting the roots in Eq.\eqref{log-free-series} gives 

\begin{align}
\phi_{i}= c^{\gamma_i} \sum\limits_{n\ge0} \frac{[\gamma_i]_{(0,n,n,0,0)}}{[\gamma_i+(n,-n,-n,0,n)]_{(n,0,0,0,n)}} 
\left(\frac{c_1 c_5}{c_2 c_3}\right)^n, 
\end{align}

\noindent which leads to the series

\begin{align}
 \phi_1=& c^{\gamma_1} \sum\limits_{n\ge0} \frac{\poch{\beta-\alpha_2-\alpha_3} \poch{\beta-\alpha_1-\alpha_3}}
 {\poch{1}\poch{\beta-A+1} } \left(\frac{c_1 c_5}{c_2 c_3}\right)^n,\\
 \phi_2= &c^{\gamma_2} \sum\limits_{n\ge0} \frac{\poch{\alpha_1} \poch{\alpha_2}}
 {\poch{-\beta+A+1} \poch{1} } \left(\frac{c_1 c_5}{c_2 c_3}\right)^n.
\end{align}

\noindent Integration constants can be easily computed by setting $c_1=0$ and $c_5=0$ in Eq.\eqref{toric-triangle}
with $\Omega=\mathbb{R}_+^3$. We write them collectively as   

\begin{align}
K_r =  \frac{1}{\Gamma(\beta)} \prod\limits_{i\ne 0} \Gamma(-\gamma_r^i).
\end{align}

\noindent Taking the limit $c_1\rightarrow0$ and setting $c_2=c_3=c_4=1$, $c_5=s$, we obtain

\begin{align}
 I(\alpha)=  \frac{\Gamma(\beta-\alpha_2-\alpha_3) \Gamma(\beta-\alpha_1-\alpha_3)\Gamma(\alpha_3)
 \Gamma(\beta-A)}{\Gamma(\beta)} s^{\beta-A}.
\end{align}

\noindent Let us remark that in this example we have set $c_1$ to zero in order to compute one of the 
integration constants--- as can be seen from the roots---therefore reaching a tautology. We can fix this by choosing an appropriate weight vector such 
that none of the roots contain zero in position $1$. This leads to a more complicated version of the canonical series which
will require analytic continuation in order to take the limit $c_1\rightarrow 0$.  Since we want to 
interpret Feynman integrals with codimension zero matrix as a certain limiting cases of
$\sfA$-hypergeometric functions, we do not worry about this situation. In fact, as we will see in the next
example, we can take advantage of this by choosing a weight vector such that all massless
cantaloupe graphs are defined by the coefficients of a linear  combination of two  Gauss hypergeometric
functions.

\subsubsection{The massless $L$-loop cantaloupe graph}
\begin{figure}[ht]
 \centering
\begin{tikzpicture}
\draw[dashed, thick] (0,0) ellipse (3cm and 2cm);
\draw[dashed, thick] (0,0) ellipse (3cm and 1.cm);
\draw [dashed, thick](0,0) ellipse (3cm and 0.5cm);
\draw (-3,0)--(-3.5,0);
\draw (3,0)--(3.5,0);
\node[text width=0.5cm, text centered ] at (0,0.1) {$\vdots$};
\node [text width=0.5cm, text centered ] at (-4,0) {$p$};
\node [text width=0.5cm, text centered ] at (0,-2.2){$1$};
\node [text width=0.5cm, text centered ] at (0,-1.2){$2$};
\node [text width=0.5cm, text centered ] at (0,1.2){$L$};
\node [text width=1.5cm, text centered ] at (0,2.2){$L+1$};
\end{tikzpicture} 
\caption{Massless cantaloupe graph}
\label{cantaloupe-diagram}
\end{figure}
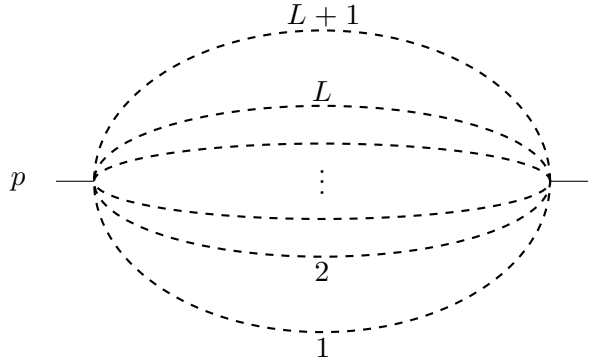

Let us consider the case of the $L$-loop cantaloupe  graph (a.k.a. banana graph) shown in 
Fig.\ref{cantaloupe-diagram}. We parametrize the inverse propagators as follows

\begin{align}
 D_1=(k_1-p)^2, \quad D_2=(k_2-k_1)^2,\dots, D_{L}=(k_{L}-k_{L-1})^2,\quad D_{L+1}=(k_{L})^2. 
\end{align}

\noindent The  polynomial $g(z)$ of this graph can be written  in general as

\begin{align}
g(z_1, \dots,z_{L+1})=\sum\limits_{i=1}^{L+1}  \prod\limits_{j\ne i}^{L} z_j+s\prod\limits_{i=1}^{L+1}{z_i},     
\end{align}
\noindent where $s=-p^2$. The integral to be computed reads  

\begin{align}
 I(\alpha)=  \int_{\mathbb{R}_+^{L+1}} \dd\eta_{L+1} \frac{z_1^{\alpha_1} \cdots z_{L+1}^{\alpha_{L+1}} }
 {g(z)^\beta}.
\end{align}

\noindent It is easy to check that the $g(z)$ polynomials of this graph lead to codimension zero matrices and
hence  they fall under the class of problems where we must introduce a deformation to define a GKZ 
system in the sense of Eq.\eqref{log-free-series}. In order to perform such deformation  systematically,
let us introduce some notation. Let $\mathsf{1}_i$ denote  a sequence of $1$'s of length $i$ and similarly
for $\mathsf{0}_j$. We have the  relation $i+j=L+1$. Furthermore, let   

\begin{align}
 v:=(1_{L-1},0_{2}).
\end{align}

\noindent At each loop, we set a deformation monomial
\begin{align}
r(z)= c_1 z^{v}, 
\end{align}

\noindent hence we have 

\begin{align}
 g_r(c, z)= c_1 z^{v} +\sum\limits_{i=1}^{L+1} c_{L+3-i} \prod\limits_{j\ne i}^{L} z_j+c_{L+3}\prod\limits_{i=1}^{L+1}{z_i},
\end{align}

\noindent where $c_{L+3}=s$. Let us give an example. For $L=3$, $v=(1,1,0,0)$ and $r(z)=c_1 z_1 z_2$, then we have the deformed
toric polynomial 

\begin{align}
 g_r(c, z)=c_1 z_1 z_2 +  c_2 z_1 z_2 z_3+ c_3 z_1 z_2 z_4+ c_4 z_1 z_3 z_4 +c_5 z_2 z_3 z_4 + c_6 z_1 z_2 z_3 z_4. 
\end{align}

\noindent After introducing the deformation, the $(L+2)\times (L+3) $ matrix associated with the $L$-loop cantaloupe 
graph can be written in the general form

\begin{align}
\sfA= \begin{pmatrix}
  1&1& \dots & & 1 &1\\
  & \mathsf{1}_{L+1}& &  &  0&\mathsf{1}_1\\
  &\mathsf{1}_{L}& &   &0&\mathsf{1}_2\\
  && &  &  &&\\
    & &  \vdots&   & &  \\
      && &  &  &&\\
   & \mathsf{1}_3&0&   &  &\mathsf{1}_{L-1}  \\
   0&\mathsf{1}&0&    & &\mathsf{1}_{L} \\
   0&0&\mathsf{1}&  &  & 1_{L}  
 \end{pmatrix}.
\end{align}

\noindent Notice that each row contains $L+3$ elements. There are $L$ rows with  have a single zero and 
$2$ rows with two zeros. The  integral under consideration is given by 

\begin{align}
 I_{g_r}(\kappa)=
 \int_{\Omega} \dd\eta_{L+1} \frac{z_1^{\alpha_1} \cdots z_{L+1}^{\alpha_{L+1}}}
 {g_r(c, z)^\beta},
\end{align}

\noindent where $\kappa=(-\beta, -\alpha_1,\dots, -\alpha_{L+1})$. Computing the  kernel of the above matrix leads to 

\begin{align}
\mathcal{L}=\mathbb{Z}(1, -1,-1, 0_{L-1}, 1),
\end{align}

\noindent where by definition $0_0:=\emptyset$.  We choose $w=(1,\mathsf{0}_{L+2})$, thus obtaining

\begin{align}
 \finw= 
 \braket{ \theta_1 \theta_{L+3}}+\braket{\mathsf{A}\theta-\kappa^T}.
\end{align}

\noindent The roots can be written as 

\begin{align}
 \{\gamma_i\}=&\{
 \left(0,\alpha_{L+1}-\beta,\ \dots, \ \alpha _1-\beta,L\beta-\sum\limits_{i=1}^{L+1}\alpha_i\right),\\
 &\left(\sum\limits_{i=1}^L\alpha_i-L \beta,\ 
 (L-1) \beta-\sum_{i=1}^{L} \alpha_i, \ (L-1) \beta-\sum_{i\ne L}^{L+1} \alpha_i,\
 -\beta+\alpha_{L-1},\ \dots,\ -\beta+\alpha _1,\ 0\right)\nonumber
 \},
\end{align}

\noindent which lead to the canonical series 

\begin{align}
 \phi_{1}=& c^{\gamma_{1}}\; {}_2F_1\left(\beta-\alpha_{L+1},\; \beta-\alpha_{L},\; 
 L\beta-\sum_{i=1}^{L+1}\alpha_i+1; x
\right),\\
 \phi_{2}=& c^{\gamma_{2}}\; {}_2F_1\left(-(L-1) \beta+\sum_{i=1}^{L} \alpha_i,\; 
 -(L-1) \beta+\sum_{i\ne L}^{L+1} \alpha_i ; \sum\limits_{i=1}^L\alpha_i-L \beta+1; x\right),
\end{align}

\noindent where

\begin{align}
 x=\frac{c_1 c_{L+3} }{c_2 c_3}
\end{align}

\noindent The relevant integration constant reads

\begin{align}
 K_1= \frac{\Gamma(-L \beta+\sum_{i=1}^{L+1}\alpha_i)}{\Gamma(\beta)} \prod\limits_{i=1}^{L+1} \Gamma(\beta-\alpha_i),
\end{align}

\noindent which after a change of variables corresponds to the formula of the Mellin transform of a linear function. 
Setting $c_1=0$, $c_2=\dots=c_{L+2}=1$,  and $c_{L+3}=s$ we arrive at

\begin{align}
I(\alpha)=  s^{\left(L\beta-\sum_{i=1}^L\alpha_i\right)} \frac{\Gamma(-L \beta+\sum_{i=1}^{L+1}\alpha_i)}{\Gamma(\beta)} 
\prod\limits_{i=1}^{L+1} \Gamma(\beta-\alpha_i).
\end{align}

\noindent  The formulas for the fake indicial ideal and its roots  have been checked up to 5-loop. In this example, we have seen 
that deforming $g(z)$ leads to a GKZ system of $\text{co}(\sfA)=1$ and taking  the limit of the deformation
to zero at the end of the computation allows us to interpret Feynman integrals as the limit of a linear
combination of their canonical series.  

\subsection{Examples of $\text{co}(\sfA)=1$}

\subsubsection{One-mass bubble}
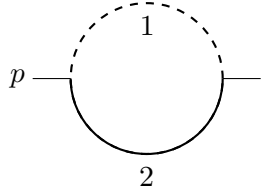
\begin{figure}[ht]
\centering
\begin{tikzpicture}
\draw [dashed, thick](0,1)circle (1);
\draw(-1.5,1)--(-1.0,1);
\draw(1.0,1)--(1.5,1);
\draw[thick] (1,1) arc (0:-180:1);
\node[text width=0.5cm, text centered ] at (0,1.7) {$1$};
\node[text width=0.5cm, text centered ] at (0,-0.3) {$2$};
\node[text width=0.5cm, text centered ] at (-1.7,1.0) {$p$};
\end{tikzpicture} 
\caption{Single mass bubble graph}
\label{onemass-bubble}
\end{figure}

Let us work now with the bubble integral with one massive internal line (Fig.\ref{onemass-bubble}). The inverse
propagators of this integral read 

\begin{align}
D_1= (k^2), \qquad D_2 =(k-p)^2+m^2 .
\end{align}

\noindent Omitting the overall $\Gamma$ factors we have the Lee-Pomeransky representation

\begin{align}
  I(\alpha)=&  \int_{\mathbb{R}^2_+} \frac{z_1^{\alpha_1} z_2^{\alpha_2} }{
 (z_1+z_2+ (m^2+s)z_1 z_2+m^2 z_2^2)^{\beta}} 
 \frac{\dd z_1}{z_1} \frac{\dd z_2}{z_2}, 
 \label{integral-onemass-bubble}
\end{align}

\noindent where $s=-p^2$. Therefore, the corresponding toric polynomial and the associated  matrix $\sfA$ read

\begin{align}
g(c, z)=c_1 z_1+ c_2 z_2+c_3 z_1 z_2+c_4 z_2^2 \Longleftrightarrow \mathsf{A}=\left(
\begin{array}{cccc}
 1 & 1 & 1 & 1 \\
 1 & 0 & 1 & 0 \\
 0 & 1 & 1 & 2 \\
\end{array}
\right).
\label{pol-one-mass-bubble}
\end{align}

\noindent  Now, let us consider the related problem from the GKZ point of view. We are interested in the
more general situation

 \begin{align}
I_g(\kappa)=&  \int_{\Omega} \frac{z_1^{\alpha_1} z_2^{\alpha_2} }{(c_1 z_1+ c_2 z_2+c_3 z_1 z_2+c_4 z_2^2)^{\beta}} 
 \frac{\dd z_1}{z_1} \frac{\dd z_2}{z_2}. 
 \label{integral-onemass-bubble-toric}
\end{align}

\noindent Taking $w=(0,1,1,1)$, we obtain the following fake indicial ideal

\begin{align}
 \finw=\braket{\theta _2 \theta _3,\beta +\theta _1+\theta _2+\theta _3+\theta _4,\alpha _1+\theta _1+\theta _3,
 \alpha _2+\theta _2+\theta _3+2 \theta _4}.
\end{align}

\noindent Then, the roots of this ideal read 

\begin{align}
\{\gamma_i\}= \{(-\alpha _1,2 \alpha _1+\alpha _2-2 \beta ,0,-\alpha _1-\alpha _2+\beta ),
(\alpha _1+\alpha _2-2 \beta ,0,-2 \alpha _1-\alpha _2+2 \beta ,\alpha _1-\beta )\}.
\end{align}

\noindent In addition, we have 
$\mathcal{L}=\mathbb{Z}(-1,1,1,-1)$, and hence $u=n(-1,1,1,-1)$. We also have  $u.w=n$, which implies $[\gamma_i]_{u_-}=0$
for $n<0$. The canonical series simplify to the functions

\begin{align}
\phi_{1}=& c^{\gamma_1}{}_2F_1\left(\alpha_1, \alpha_1+\alpha_2-\beta;
2\alpha_1+\alpha_2-2\beta+1;\frac{c_2 c_3}{c_1c_4}\right),\\
\phi_{2}=& c^{\gamma_2}{}_2F_1\left(2\beta-\alpha_1-\alpha_2,
\beta-\alpha_1;2\beta-2\alpha_1-\alpha_2+1;\frac{c_2 c_3}{c_1c_4}\right).
\end{align}

\noindent The result of the integral is a linear combination of $\phi_{{1,2}}$. Constants of integration
can be obtained by integrating \eqref{integral-onemass-bubble-toric} taking $c_3=0$ and 
$\Omega=\mathbb{R}_+^2$, and similarly for $c_2$. This leads to the multiplying factors 

\begin{align}
 K_1=&\frac{\Gamma(\alpha_1)\Gamma(\alpha_2+\alpha_1-\beta)\Gamma(2\beta-2\alpha_1-\alpha_2)}
{\Gamma(\beta)},\\
K_2=& \frac{\Gamma(\beta-\alpha_1)
\Gamma(2\alpha_1+\alpha_2 -2\beta)\Gamma( 2\beta- \alpha_1-\alpha_2)}{\Gamma(\beta)}.
\end{align}

\noindent Setting $c_3=(s+m^2)$  and $c_4=m^2$, the resulting integral reads

\begin{align}
 I(\alpha)=&  (m^2)^{\beta-\alpha_1-\alpha_2}\Big( K_1  \;
 {}_2F_1\left(\alpha_1, \alpha_1+\alpha_2-\beta;
2\alpha_1+\alpha_2-2\beta+1;1+s/m^2\right)\\
  &+  (1+s/m^2)^{2\beta-2\alpha_1-\alpha_2}K_2\;
{}_2F_1\left(2\beta-\alpha_1-\alpha_2,
\beta-\alpha_1;2\beta-2\alpha_1-\alpha_2+1;1+s/m^2\right)\Big).
\nonumber
\end{align}

\noindent The form of the result as a sum of two hypergeometric functions is reminiscent of the negative dimension 
approach \cite{Anastasiou:1999ui}.  Using the Eq.\eqref{id4}, we obtain

\begin{align}
 I(\alpha)=&
 (m^2)^{\beta-\alpha_1-\alpha_2} \frac{\Gamma \left(\alpha _1\right) \Gamma \left(\beta -\alpha _1\right)
 \Gamma \left(2 \beta -\alpha _1-\alpha _2\right) \Gamma \left(-\beta +\alpha _1+\alpha _2\right)}{\Gamma (\beta )^2}\\
& \qquad \times {}_2F_1\left(\alpha_1,-\beta+\alpha_1+\alpha_2;\beta; -s/m^2\right). \nonumber
\end{align}

\subsubsection{One-mass sunset}
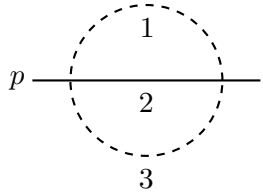
\begin{figure}[ht]
\centering
\begin{tikzpicture}
%\draw[step=1cm,black,very thin] (-1.9,-1.9) grid (1.9,1.9);
\draw [dashed, thick] (0,1)circle (1);
\draw[thick](-1.5,1)--(1.5,1);
\node[text width=0.5cm, text centered ] at (0,0.7) {$2$};
\node[text width=0.5cm, text centered ] at (0,1.7) {$1$};
\node[text width=0.5cm, text centered ] at (0,-0.3) {$3$};
\node[text width=0.5cm, text centered ] at (-1.7,1.0) {$p$};
\end{tikzpicture} 
\caption{Sunset graph}
\label{sunrise}
\end{figure}

We now consider the single  mass sunset graph with the constraint $s=-p^2=m^2$ (Fig.\ref{sunrise}). The 
inverse  propagators read

\begin{align}
 D_1=(k_1-p_1)^2, \qquad  D_2=(k_2-k_1)^2+m^2 ,\qquad D_3= k_2^2,
\end{align}
\noindent which lead to the integral 
\begin{align}
I(\alpha)= \int\limits_{\mathbb{R}_+^3} \dd\eta_3 \frac{z_1^{\alpha_1} z_2^{\alpha_2} z_3^{\alpha_3}}{
(z_1z_2+ z_1 z_3 + z_2 z_3 + m^2 z_1 z_2^2+ m^2 z_2^2 z_3)^\beta}. 
\end{align}
\noindent We consider the following toric polynomial
\begin{align}
g(c, z)=c_1 z_1z_2+c_2 z_1 z_3 + c_3 z_2 z_3 + c_4 z_1 z_2^2+ c_5 z_2^2 z_3 \Longleftrightarrow \sfA=\left(
\begin{array}{ccccc}
 1 & 1 & 1 & 1 & 1 \\
 1 & 1 & 0 & 1 & 0 \\
 1 & 0 & 1 & 2 & 2 \\
 0 & 1 & 1 & 0 & 1 \\
\end{array}
\right),
\end{align}

\noindent such that the integral under consideration becomes 

\begin{align}
I_g(\kappa)= \int\limits_{\Omega} \dd\eta_3 \frac{z_1^{\alpha_1} z_2^{\alpha_2} z_3^{\alpha_3}}{
(c_1 z_1z_2+ c_2 z_1 z_3 + c_3 z_2 z_3 + c_4 z_1 z_2^2+ c_5 z_2^2 z_3)^\beta}. 
\end{align}

\noindent Choosing the weight vector $w=(0,1,1,1,1)$, we find 

\begin{align}
\finw=&\braket{\theta _3 \theta _4}+\braket{\sfA \theta-\kappa^T},\\
 \{\gamma_i\}=&\{ (\alpha _3-\beta ,-\alpha _1-\alpha _3+\beta ,2 \alpha _1+\alpha _2+\alpha _3-3 \beta ,0,-\alpha _1-\alpha _2-\alpha _3+2 \beta ),\\
  & (2 \alpha _1+\alpha _2+2 \alpha _3-4 \beta ,-\alpha _1-\alpha _3+\beta ,0,-2 \alpha _1-\alpha _2-\alpha _3+3 \beta ,\alpha _1-\beta ) \}.
  \nonumber
\end{align}

\noindent From $\mathcal{L}= \mathbb{Z}(1,0,-1,-1,1)$, we have $u=n(1,0,-1,-1,1)$, hence $u.w=-n$. In order to start the 
sum from $n=0$, we set $u\rightarrow-u$ and thus $u=n(0,0,1,1,0)-n(1,0,0,0,1)$. We obtain

\begin{align}
 \phi_{1}=&c^{\gamma_1} \sum\limits_{n\ge0}
 \frac{\poch{\beta -\alpha _3} \poch{\alpha _1+\alpha _2+\alpha _3-2 \beta}}
 {\poch{2 \alpha _1+\alpha _2+\alpha _3-3 \beta +1} \poch{1}}\left(\frac{c_3 c_4}{c_1 c_5}\right)^n,\\ 
  \phi_{2}=&c^{\gamma_2} \sum\limits_{n\ge0}
 \frac{\poch{-2 \alpha _1-\alpha _2-2 \alpha _3+4 \beta} \poch{\beta -\alpha _1}}
 {\poch{1} \poch{-2 \alpha _1-\alpha _2-\alpha _3+3 \beta +1}}\left(\frac{c_3 c_4}{c_1 c_5}\right)^n.
\end{align}

\noindent The integration constants can be written collectively as

\begin{align}
 K_r=&\frac{1}{\Gamma(\beta)} \prod\limits_{i\ne 0} \Gamma(-\gamma_r^i).
\end{align}

\noindent Finally, setting $c_1=c_2=c_3=1$ and $c_4=c_5=m^2$, we arrive at the result

\begin{align}
I(\alpha)=& (m^2)^{-\alpha_1-\alpha_2-\alpha_3+2\beta}\Big(K_1  \; {}_2F_1(\beta -\alpha _3, \alpha-2 \beta; \alpha+\alpha_1-3 \beta +1; 1)
\label{onemass-sunset-final}\\
&+K_2 \; {}_2F_1( -2\alpha+\alpha_2+4\beta,  \beta-\alpha_1; -\alpha-\alpha_1+3 \beta +1; 1)\Big).\nonumber
\end{align}

\noindent In order to write this result in simpler form let 

\begin{align}
 A=\beta-\alpha_3,\quad B= -2\beta+ \alpha _1+\alpha _2+\alpha _3, \quad C=2\beta-\alpha _1-\alpha _3,\nonumber
\end{align}

\noindent hence we can write Eq.\eqref{onemass-sunset-final} as 

\begin{align}
I(\alpha)= (m^2)^{-B}
\frac{\Gamma (A) \Gamma (B) \Gamma \left(\beta-C\right) \Gamma (C-A) \Gamma (C-B)}{\Gamma\left(C\right)\Gamma
\left(\beta\right)},
\end{align}
\noindent where we have used identity \eqref{id4}.

\subsubsection{Single scale party hat}
The next example is the party hat graph shown in Fig.\ref{party-hat}. We have the inverse propagators

\begin{align}
 D_1=k_2^2, \quad D_2=(k_2-p_3)^2,\quad D_3=(k_2-k_1)^2,\quad D_4=(k_1-p_1)^2.
\end{align}

\noindent We will consider the case $-p_1^2=-p_3^2=0$ and $s=p_2^2$. The relevant integral in the Lee-Pomeransky
representation reads 

\begin{align}
  I(\alpha)=&  \int_{\mathbb{R}^4_+} \dd\eta_4 \frac{z_1^{\alpha_1} z_2^{\alpha_2} z_3^{\alpha_3} z_4^{\alpha_4}}{
 (z_1 z_3+z_1 z_4+z_2 z_3+z_2 z_4+z_4 z_3+ s z_2 z_4 z_3)^{\beta}}. 
\end{align}

\noindent Therefore, the toric polynomial has the form
\begin{align}
 g(c,z)=c_1 z_1 z_3+ c_2 z_1 z_4+ c_3z_2 z_3+ c_4 z_2 z_4+c_5 z_3 z_4 + c_6 z_2 z_3  z_4,
\end{align}

\noindent where $c_6=s$. We associate the following matrix to $g(c,z)$ 

\begin{figure}[ht]
\centering
\begin{tikzpicture}
%\draw[step=1cm,black,very thin] (-1.9,-1.9) grid (1.9,1.9);
\draw[dashed, thick](0,-1) -- (1, 1);
\draw[dashed, thick](0,-1) -- (-1, 1);
\draw[dashed, thick](-1,1) -- (1, 1) arc (-1:180:1);
\draw[dashed, thick](-1,1)--(-1.5,1.5);
\draw[thick](1,1)--(1.5,1.5);
\draw[dashed](0,-1)--(-0.0,-1.5);
\node[text width=0.5cm, text centered ] at (-1,0) {$1$};
\node[text width=0.5cm, text centered ] at (1,0) {$2$};
\node[text width=0.5cm, text centered ] at (0,0.7) {$3$};
\node[text width=0.5cm, text centered ] at (0,1.7) {$4$};
\node[text width=0.5cm, text centered ] at (0,1.7) {$4$};
\node[text width=0.5cm, text centered ] at (1.7,1.0) {$p_2$};
\end{tikzpicture}
\caption{Party hat}
\label{party-hat}
\end{figure}

\begin{align}
\sfA= \left(
\begin{array}{cccccc}
 1 & 1 & 1 & 1 & 1 & 1 \\
 1 & 1 & 0 & 0 & 0 & 0 \\
 0 & 0 & 1 & 1 & 0 & 1 \\
 1 & 0 & 1 & 0 & 1 & 1 \\
 0 & 1 & 0 & 1 & 1 & 1 \\
\end{array}
\right).
\label{party-hat-matrix}
\end{align}

\noindent Taking  $w=(0,1,1,1,1,1)$ leads to the fake indicial ideal (see example Appendix \ref{glossary})

\begin{align}
 \finw=& \langle \theta _2 \theta _3,
 \beta +\theta _1+\theta _2+\theta _3+\theta _4+\theta _5+\theta _6,
 \alpha _1+\theta _1+\theta _2,\alpha _2+\theta _3+\theta _4+\theta _6,\\ 
& \qquad \alpha _3+\theta _1+\theta _3+\theta _5+\theta _6,\alpha _4+\theta _2+\theta _4+\theta _5+\theta _6
\rangle,\nonumber
\end{align}
\noindent with roots

\begin{align}
\gamma_i=\{&(-\alpha _1,0,\alpha _1+\alpha _4-\beta ,\alpha _3-\beta ,\alpha _1+\alpha _2-\beta ,-\alpha _1-\alpha _2-\alpha _3-\alpha _4+2 \beta),\\
&(\alpha _4-\beta ,-\alpha _1-\alpha _4+\beta ,0,\alpha _1+\alpha _3+\alpha _4-2 \beta ,\alpha _1+\alpha _2-\beta ,-\alpha _1-\alpha _2-\alpha _3-\alpha _4+2 \beta)\}.
\nonumber
\end{align}

\noindent  Computing  $\ker(\sfA)$ leads to 

\begin{align}
 \mathcal{L}=\mathbb{Z}(1,-1,-1,1,0,0) \Rightarrow u=n(1,0,0,1,0,0)-n(0,1,1,0,0,0).
\end{align}

\noindent Since  $u.w=-n$, $[\gamma]_{u_-}=0$ for $n>0$. Then, the canonical series \eqref{log-free-series} 
leads to

\begin{align}
\phi_{1}=&c^{\gamma_1} {}_2F_1(\alpha_1,\beta-\alpha_3;\alpha _1+\alpha _4-\beta +1; 
(c_1 c_4)/(c_2c_5)), \\
\phi_{2}=&c^{\gamma_2} {}_2F_1(\beta-\alpha_4,-\alpha _1-\alpha _3-\alpha _4+2 \beta;
-\alpha _1-\alpha _4+\beta +1; (c_2 c_4)/(c_1c_5)).
\end{align}

\noindent Integration constants read 

\begin{align}
 K_r= \frac{1}{\Gamma(\beta)}\prod\limits_{i\ne0}\Gamma(-\gamma_{r}^i).
\end{align}

\noindent Setting $c_1=c_2=c_3=c_4=c_5=1$, $c_6=s$, and defining 

\begin{align}
 A=\beta-\alpha_4 ,\quad B=2\beta-\alpha_1-\alpha_3-\alpha_4, \quad C=2\beta-\alpha_3-\alpha_4,
\end{align}

\noindent we arrive at

\begin{align}
I(\alpha)= s^{B-\alpha_2} 
\frac{\Gamma (A) \Gamma (B) \Gamma (C-A) \Gamma \left(\alpha _2-B\right) \Gamma (C-B)
\Gamma \left(B-C+\beta -\alpha _2\right)}{\Gamma(C)\Gamma (\beta )},
\end{align}

\noindent where we have used identity \eqref{id4}.

\subsubsection{On-shell massless box}

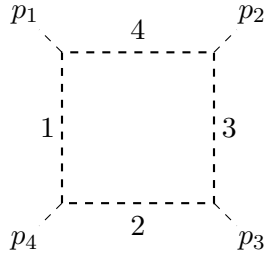
\begin{figure}[ht]
\centering
\begin{tikzpicture}
\draw[dashed](-1,-1)--(-1.3,-1.3);
\draw[dashed](1,-1)--(1.3,-1.3);
\draw[dashed](1,1)--(1.3,1.3);
\draw[dashed](-1,1)--(-1.3,1.3);
\draw[dashed, thick](-1,-1)--(-1.0,1.0)--(1.0,1.0)--(1.0,-1.0)--cycle;
\node[text width=0.5cm, text centered ] at (-1.2,0) {$1$};
\node[text width=0.5cm, text centered ] at (0,-1.3) {$2$};
\node[text width=0.5cm, text centered ] at (1.2, 0) {$3$};
\node[text width=0.5cm, text centered ] at (0,1.3) {$4$};
\node[text width=0.5cm, text centered ] at (-1.5,1.5){$p_1$};
\node[text width=0.5cm, text centered ] at (1.5,1.5){$p_2$};
\node[text width=0.5cm, text centered ] at (1.5,-1.5){$p_3$};
\node[text width=0.5cm, text centered ] at (-1.5,-1.5){$p_4$};
\end{tikzpicture} 
\caption{On shell massless box}
\label{os-massless-box}
\end{figure}

Let us now consider the massless box integral shown in Fig.\ref{os-massless-box}. The 
inverse propagators are given by 
\begin{align}
 D_1=(k-p_1)^2, \qquad  D_2= (k+p_2+p_3)^2,\qquad  D_3=(k+p_2)^2\qquad D_4= k_1^2, 
\end{align}

\noindent where $-p_i^2=0$, $i=1,\dots,4$. We obtain   

\begin{align}
 g(z)= z_1+z_2+z_3+z_4 +s z_1 z_3 + t z_2 z_4,
\end{align}

\noindent where $s=-(p_1+p_2)^2$ and $t=-(p_2+p_3)^2$ are the usual Mandelstam invariants. From this 
polynomial we  obtain the matrix 

\begin{align}
\mathsf{A}=\left(
\begin{array}{cccccc}
 1 & 1 & 1 & 1 & 1 & 1 \\
 1 & 0 & 0 & 0 & 1 & 0 \\
 0 & 1 & 0 & 0 & 0 & 1 \\
 0 & 0 & 1 & 0 & 1 & 0 \\
 0 & 0 & 0 & 1 & 0 & 1 \\
\end{array}
\right).
\end{align}

\noindent Let us consider the more general problem

 \begin{align}
I_g(\kappa)=&  \int_{\Omega} \frac{z_1^{\alpha_1} z_2^{\alpha_2} z_3^{\alpha_3} z_4^{\alpha_4}}
{(c_1 z_1+c_2 z_2+c_3 z_3+c_4 z_4+c_5 z_3 z_1+c_6 z_2 z_4)^\beta} \dd \eta_4.
 \label{integral-box}
\end{align}

\noindent Choosing $w=(0, 1, 0, 0, 0, 0)$, the fake indicial ideal reads

\begin{align}
 \finw = \braket{\theta _2 \theta _4 \theta _5}+ \braket{\sfA\theta-\kappa^T},
\end{align}

\noindent which lead to the roots
\begin{align}
 \{\gamma_i\}=&\{( -\alpha _1,\alpha _1+\alpha _3+\alpha _4-\beta ,-\alpha _3,\alpha _1+\alpha _2+\alpha _3-\beta ,0,-\alpha _1-\alpha _2-\alpha _3-\alpha _4+\beta),\nonumber\\
  &(\alpha _2+\alpha _3-\beta ,\alpha _4-\alpha _2,\alpha _1+\alpha _2-\beta ,0,-\alpha _1-\alpha _2-\alpha _3+\beta ,-\alpha _4), \nonumber\\
  &(\alpha _3+\alpha _4-\beta ,0,\alpha _1+\alpha _4-\beta ,\alpha _2-\alpha _4,-\alpha _1-\alpha _3-\alpha _4+\beta ,-\alpha _2)\}
\end{align}
\noindent Computing $\ker \sfA$ leads to  $u= n(-1, 1, -1, 1, 1, -1)$, hence $u.w=n$. Therefore
$[\gamma]_{u_-}=0$ for $n<0$. We also  have 

\begin{align}
 u_{-}=(n, 0, n, 0, 0, n),\qquad  u_{+}=(0, n, 0, n, n, 0).
\end{align}

\noindent We can now write the canonical series \eqref{log-free-series} as

\begin{align}
 \phi_{1}=& c^{\gamma_1} \sum\limits_{n\ge0} \frac{\left(\alpha _1\right)_n \left(\alpha _3\right)_n \left(-\beta +\alpha _1+\alpha _2+\alpha _3+\alpha _4\right)_n}
 {(1)_n \left(-\beta +\alpha _1+\alpha _2+\alpha _3+1\right)_n \left(-\beta +\alpha _1+\alpha _3+\alpha _4+1\right)_n}
 \left(\frac{c_2 c_4 c_5}{c_1 c_3 c_6}\right)^n,\\
 \phi_{2}=& c^{\gamma_2} \sum\limits_{n\ge0} \frac{\left(\alpha _4\right)_n \left(\beta -\alpha _1-\alpha _2\right)_n \left(\beta -\alpha _2-\alpha _3\right)_n}
 {(1)_n \left(-\alpha _2+\alpha _4+1\right)_n \left(\beta -\alpha _1-\alpha _2-\alpha _3+1\right)_n}
  \left(\frac{c_2 c_4 c_5}{c_1 c_3 c_6}\right)^n,\\
  \phi_{3}=& c^{\gamma_3} \sum\limits_{n\ge0}\frac{\left(\alpha _2\right)_n \left(\beta -\alpha _1-\alpha _4\right)_n \left(\beta -\alpha _3-\alpha _4\right)_n}
  {(1)_n \left(\alpha _2-\alpha _4+1\right)_n \left(\beta -\alpha _1-\alpha _3-\alpha _4+1\right)_n}
    \left(\frac{c_2 c_4 c_5}{c_1 c_3 c_6}\right)^n.
\end{align}
\noindent The constants of integration read 

\begin{align}
K_r = \frac{1}{\Gamma(\beta)}\prod\limits_{i\ne0} \Gamma(-\gamma_r^i).    
\end{align} 

\noindent In this case the canonical series evaluate to hypergeometric functions
${}_3F_2(a,b;c;d,e;x)$ (see definition in Eq.\eqref{id3F2}). Let
$A= \alpha_1+ \alpha_2+ \alpha_3+\alpha_4$ and $x=(c_2 c_4 c_5)/(c_1 c_3 c_6)$. We can write the result in the 
condensed form

\begin{align}
 I_b(\kappa, c) &= K_1 c^{\gamma_1} {}_3F_2\left(\alpha _1,\alpha _3,-\beta +A;
 -\beta +A-\alpha_4+1,-\beta +A-\alpha_2+1;x\right) \\ 
 &+   K_2 c^{\gamma_2} \, _3F_2\left(\beta-\alpha _1-\alpha _2,\beta -\alpha _2-\alpha _3,\alpha _4;
 \beta -A+\alpha_4+1,-\alpha _2+\alpha _4+1;x\right)\nonumber\\
 &+K_3 c^{\gamma_3} {}_3F_2\left(\alpha _2,\beta -\alpha _1-\alpha _4,\beta -\alpha _3-\alpha _4;
 \alpha _2-\alpha _4+1,\beta -A+\alpha_2+1;x\right)
 \nonumber
\end{align}

\noindent Setting the constant to the values of the original polynomial, i.e., $c_1=\dots=c_4=1$, $c_5=s$ and $c_6=t$, we 
obtain

\begin{align}
I(\alpha)&=K_1 t^{\beta -A} {}_3F_2\left(\alpha _1,\alpha _3,-\beta +A;
 -\beta +A-\alpha_4+1,-\beta +A-\alpha_2+1;s/t\right) \\ 
 &+K_2 t^{-\alpha _4} s^{\alpha _4-A+\beta}
 \, _3F_2\left(\beta-\alpha _1-\alpha _2,\beta -\alpha _2-\alpha _3,\alpha _4;
 \beta -A+\alpha_4+1,-\alpha _2+\alpha _4+1;s/t\right)\nonumber\\
 &+K_3 t^{-\alpha _2} s^{\alpha _2-A+\beta }
 {}_3F_2\left(\alpha _2,\beta -\alpha _1-\alpha _4,\beta -\alpha _3-\alpha _4;
 \alpha _2-\alpha _4+1,\beta -A+\alpha_2+1;s/t\right)\nonumber.
\end{align}

\noindent Hence, we have recovered the result obtained by the  differential reduction method using Gr\"obner bases 
given  in Ref.\cite{Tarasov:2017jen} based on \cite{Tarasov:1996bz, Tarasov:1998nx}. Interestingly, the computation
of integrations constants in Ref.\cite{Tarasov:2017jen} follows a similar prescription as ours. However, in our 
case, because of our choice of $w$, we only set $c_5=0$ to compute $K_1$ but we do not use $c_6=0$ as 
in \cite{Tarasov:2017jen}. This results also matches nicely the results obtained in 
Ref.\cite{Anastasiou:1999cx} using the negative dimension approach.

\subsection{A $\text{co}(\sfA)=2$ example}

\begin{figure}
\centering
 \begin{tikzpicture}
%\draw[step=1cm,black,very thin] (-1.9,-1.9) grid (1.9,1.9);
\draw [dashed](0,-1) -- (1, 1);
\draw [dashed](0,-1) -- (-1, 1);
\draw [dashed](-1,1) -- (1, 1);
\draw(-1,1)--(-1.5,1.5);
\draw(1,1)--(1.5,1.5);
\draw(0,-1)--(-0.0,-1.5);
\node[text width=0.5cm, text centered ] at (-1,0) {$1$};
\node[text width=0.5cm, text centered ] at (1,0) {$2$};
\node[text width=0.5cm, text centered ] at (0,0.7) {$3$};
\end{tikzpicture}
\caption{Three-scale triangle: $s_1=-p_1^2$, $s_2=-p_2^2$, $s_3=-(p_1+p_2)^2$. }
\label{three-scale-triangle}
\end{figure}
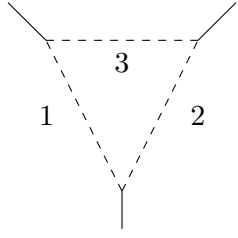

Finally, let us consider the triangle integral shown in Fig.\ref{three-scale-triangle} where all external momenta are 
nonvanishing. The inverse propagators read

\begin{align}
D_1=(k_1 - p_1)^2,\qquad D_2=(k_1+p_2),\qquad D_3=k_1^2, 
\end{align}

\noindent where we have  $-p_1^2=s_1$, $-p_2^2=s_2$,  and $-(p_1+p_2)^2=s_3$. Using momentum conservation, we have 

\begin{align}
g(z)= z_1+z_2+z_3+s_3 z_1 z_2 +s_1 z_1 z_3 + s_2 z_2 z_3 .
\end{align}

\noindent The associated matrix reads

\begin{align}
\mathsf{A}= \begin{pmatrix}
 1 & 1 & 1 & 1 & 1 & 1 \\
 1 & 0 & 0 & 1 & 1 & 0 \\
 0 & 1 & 0 & 1 & 0 & 1 \\
 0 & 0 & 1 & 0 & 1 & 1 \\
\end{pmatrix}
\end{align}

\noindent This matrix has $\text{co}(\sfA)=2$ and hence the resulting $\sfA$-hypergeometric functions will be functions of 
two variables. The integral under consideration then reads 

\begin{align}
 I_{g}(\kappa)=
 \int_{\Omega} \dd\eta_3 \frac{z_1^{\alpha_1} z_2^{\alpha_2} z_3^{\alpha_3}}{(c_1 z_1+ c_2 z_2+ c_3 z_3+ c_4 z_1 z_2 + c_5 z_1 z_3 +c_6 
 z_2 z_3)^\beta}.
 \label{toric-triangle-two-scale}
\end{align}

\noindent Computing $\ker (\sfA)$, we have 

\begin{align}
 \mathcal{L}= \{(-1,0,1,1,0,-1),(-1,1,0,0,1,-1)\},
\end{align}

\noindent which means that in this case the lattice is generated by 

\begin{align}
 u=m(-1,0,1,1,0,-1)+n(-1,1,0,0,1,-1), \quad m,n \in \mathbb{Z}. 
\end{align}

\noindent We choose $w=(0, 0, 1, 0, 0, 0)$, which has the advantage of restricting the sum over $m$. We have $u.w=m$, hence 
$[\gamma_i]_{u_-}= 0$ for $ m<0$. The fake indicial ideal reads 
\begin{align}
\finw=& \braket{\theta _2 \theta _5,\theta _3 \theta _4} + \braket{\sfA \theta -\kappa^T},
\end{align}

\noindent with roots
\begin{align}
 \{\gamma_i\} =\{ &(-\alpha _1,C-\beta ,B-\beta ,0,0,\beta -A),
 (\alpha _2-\beta ,C-\beta ,0,\beta -B,0,-\alpha _3), \\ 
 &(\alpha _3-\beta ,0,B-\beta ,0,\beta -C,-\alpha _2),
 (A-2 \beta ,0,0,\beta -B,\beta -C,\alpha _1-\beta) \} \nonumber
\end{align}

\noindent where $A=\alpha_{1}+\alpha_2+\alpha_3$, $B=\alpha_{1}+\alpha_2$, and $C=\alpha_1+\alpha_3$. Inserting the roots in 
Eq.\eqref{log-free-series} gives four solutions

\begin{align}
\phi_{1}=& c^{\gamma_1}\sum\limits_{m\ge0, n\in \mathbb{Z}} \frac{\left(\alpha _1\right)_{m+n} 
(A-\beta )_{m+n}}{(-\beta +C+1)_n (-\beta +B+1)_m(1)_m(1)_n} x^ m y^n, \\
\phi_{2}=&  c^{\gamma_2}\sum\limits_{m\ge0, n\in \mathbb{Z}}
\frac{ \left(\beta -\alpha _2\right)_{m+n} \left(\alpha _3\right)_{m+n}}
{(-\beta +C+1)_n(1)_m(\beta -B+1)_m(1)_n} x^ m y^n, \\
\phi_{3}=& c^{\gamma_3}\sum\limits_{m\ge0, n\in \mathbb{Z}} 
\frac{\left(\beta -\alpha _3\right)_{m+n} \left(\alpha _2\right)_{m+n}}
{(1)_n(-\beta +B+1)_m(1)_m(\beta -C+1)_n} x^ m y^n, \\
\phi_{4}=& c^{\gamma_4}\sum\limits_{m\ge0, n\in \mathbb{Z}} 
\frac{(2 \beta -A)_{m+n}\left(\beta -\alpha _1\right)_{m+n}}
{(1)_n(1)_m(\beta -B+1)_m(\beta -C+1)_n} x^ m y^n,
\end{align}

\noindent where we have defined $x=(c_3 c_4)/(c_1 c_6)$ and $y=(c_2 c_5)/(c_1 c_6)$. Due to our choice 
of weight vector, the sum has been only restricted in $m$. However, it is easy to see that the sum over
negative integers $n$ vanish due to the presence of $(1)_n$. Hence the above
sums have the form

\begin{align}
 \sum\limits_{m,n\ge0 } \frac{\left(a\right)_{m+n} \left( b \right)_{m+n}}{\left( c\right)_{m}
 \left(d\right)_n (1)_m (1_n)} x^m  y^n
\end{align}

\noindent which are sum representations of the Appell hypergeometric function $F_4$(See Eq.\eqref{idF4}). In order to compute the integration constants we 
can follow our prescription and set $c_4=c_5=0$ in
Eq.\eqref{toric-triangle-two-scale} and take $\Omega=\mathbb{R}^3$. This computes the first coefficient in the 
expansion $K_1$.  We follow the same procedure for the remaining integration constants. They can can collectively
be written as

\begin{align}
 K_r = \frac{1}{\Gamma(\beta)} \sum\limits_{i \ne 0} \Gamma(-\gamma_r^i).
\end{align}

\noindent Setting $c_1=c_2=c_3=1$ and $c_4=s_3$, $c_5=s_1$, and $c_6=s_2$, we can write the Feynman integral
as

\begin{align}
 I(\alpha)=& K_1 s_2^{\beta-A} 
  F_4(\alpha_1, A-\beta; -\beta+\alpha_{13}+1, -\beta+\alpha_{12}+1; s_3/s_2, s_1/s_2) \\
  &+ K_2 s_2^{-\alpha _3} s_3^{\beta -B} F_4(\beta-\alpha_2, \alpha_3; C-\beta +1,
  -B+\beta +1; s_3/s_2, s_1/s_2) \nonumber \\
 & + K_3 s_2^{-\alpha _2} s_1^{\beta -C} F_4(\beta-\alpha_3, \alpha_2; B-\beta +1,
  -C+\beta +1;  s_3/s_2, s_1/s_2)\nonumber \\
&+ K_4 s_1^{\beta -C} s_2^{\alpha _1-\beta } s_3^{\beta -B}
F_4(2 \beta -A, \beta -\alpha _1; -B+\beta +1, 
-C+\beta +1;  s_3/s_2, s_1/s_2), \nonumber
\end{align}

\noindent which agrees with the results obtained via the Mellin-Barnes integral representations
\cite{Boos:1987bg, Boos:1990rg} and the negative dimension approach \cite{Anastasiou:1999ui}.

\section{Conclusions and outlook}
\label{conclusions}
In this paper have studied the relation between the Lee-Pomeransky representation of Feynman integrals and GKZ systems. We have shown that in generic cases
we can associate a matrix $\sfA$ of $\text{co}(\sfA)>0$ to a deformed polynomial $g_r(c, z)=r(c, z)+\mathcal{U}(c)+\mathcal{F}(c)$, 
where $r(c, z)$ is introduced to ensure a canonical series representation.  
$\mathcal{U}(c)$ and $\mathcal{F}(c)$ are toric polynomials associated with the Symanzik polynomials.
Under these restrictions, we can interpret a large class of Feynman integrals as furnishing a solution of 
a GKZ system based on $\sfA$. The canonical series algorithm then allows us to evaluate integrals with arbitrary powers in the 
propagators as linear combinations of $\sfA$-hypergeometric functions. Feynman integrals are recovered at the end of the 
computation by identifying the coefficients of the toric polynomials with their kinematic values. 

Using the  canonical series method, we have evaluated several integrals for arbitrary noninteger powers in the propagators. A 
particularly nontrivial example is the on-shell massless box with arbitrary powers in the 
propagators. With this method the result was obtained as a particular case of an $\sfA$-hypergeometric integral and it matches 
the result  based on the recurrence relations method based on Gr\"obner bases
\cite{Tarasov:2017jen} and the negative dimension approach \cite{Anastasiou:1999cx}. Another nontrivial example is the $\text{co}(\sfA)=2$ 
example of the three-scale massless triangle, which is in  agreement with the
negative dimension approach as well. It would be interesting to study the relation between those methods and the canonical series algorithm.

Computing canonical series is a straightforward computational algebra problem. However, 
these series heavily depend on the choice of a weight vector $w$,  which sets the initial ideal and effectively chooses a domain of 
convergence. This choice is tied with the available information about the integration cycle and ultimately with our ability to 
compute integration constants. Our recipe  of setting $\Omega=\mathbb{R}_+^N$ is the obvious choice and was motivated by the 
cycles,  which one obtains by studying the coamoeba of $g_r(z)$ in Euclidean kinematics. A full characterization of the coamoeba of
$g_r(z)$ might be necessary when non-Euclidean kinematics is considered and going to higher codimensions. 

The rank of the system arising from the toric version of  $g_r(z)$ is bounded by $\text{vol}(\sfA)$ which, in general, is greater than
the number of master integrals arising from IBP identities or from the Euler characteristic \cite{Bitoun:2017nre}. 
The canonical series algorithm produces $\text{vol}(\sfA)$ hypergeometric series in $\text{co}(\sfA)$ variables. They collapse to 
simpler expressions once we set the coefficients $c$ to their kinematic values. This typically involves setting one or more
of these variables to unity, which amounts to evaluate hypergeometric functions at singular points. At $\text{co}(\sfA)=1$ the functions
appearing at those limits are $\Gamma$-functions. At $\text{co}(\sfA)=2$, those limits lead to hypergeometric functions of one
variable, which then collapse to simpler expressions. It would be interesting to study the mechanism which relates the number of 
master integrals and the number of canonical series solutions.

We believe that the application of GKZ systems and canonical series to Feynman integrals is not limited to the Lee-Pomeransky
representation. Indeed, it would be interesting to apply these ideas in representations where an algebraic definition of the integration 
cycles is available. For instance, this is the case of the representation due to Baikov \cite{Baikov:1996iu}, which has
recently been studied specially in the context of  maximal cuts \cite{Harley:2017qut, Frellesvig:2017aai,
Frellesvig:2019kgj, Bosma:2017ens,Mastrolia:2018uzb}. This approach is closely related to Ref.\cite{2012arXiv1212.6103H}, where bases 
of Pfaffian systems for GKZ systems are constructed using twisted cohomology groups.     

We leave these explorations for future work.

\addsec{Acknowledgements}

We thank Einan Gardi for useful discussions and encouragement during the realization of this project. We
thank Jens Forsg{\aa}rd for providing his \textsc{Mathematica} notebook for drawing coamoebas. We 
thank Claude Duhr for useful comments. We thank the anonymous referee for useful comments. We thank the Galileo Galilei Institute
in Florence for hospitality and partial financial support for the workshop ``Amplitudes in the LHC era'' in Autumn 2018. We thank
the ETH Institute for Theoretical Studies in the framework of the program ``Periods, modular forms and scattering amplitudes''
for hospitality and partial financial support. The author's research is supported by the STFC Consolidated Grant 'Particle Physics
at the Higgs Centre'.

\appendix

\section{Useful formulas}
\label{identities}

\section*{Integrals}
\begin{align}
\int_{\mathbb{R}_+}\frac{z^{\alpha}}{(a+bz)^\beta} 
\frac{dz}{z}=& \frac{\Gamma(\beta-\alpha) \Gamma(\alpha) }{\Gamma(\beta)} a^{\alpha-\beta}b^{-\alpha},
\label{gamma-integral}\\
&\qquad \text{Re}(\alpha)>0,\qquad \text{Re}(\beta-\alpha)>0. \nonumber\\
\int_{\mathbb{R}_+} \frac{z^{\alpha}}{(1+z)^{\beta_1} (1+cz)^{\beta_2}}\frac{dz}{z}
=&\frac{\Gamma(\beta_1+\beta_2-\alpha) \Gamma(\alpha)}{\Gamma(\beta_1+\beta_2)} 
{}_2F_1( \alpha, \beta_2 ;\beta_1+\beta_2;1-c),\\
&\qquad \text{Re}(\beta_1+\beta_2)>\text{Re}(\alpha)>0,\qquad|\text{arg } c|<\pi\nonumber
\end{align}

\section*{Sum representations}

\begin{align}
 {}_2F_1(a,b,c;x)=&\sum\limits_{n\ge0} \frac{\poch{a} \poch{b}}{\poch{c}\poch{1}} x^n  \label{2f1-series-sum},\\
 {}_3F_2(a,b,c;d,e;x)=& \sum\limits_{n\ge0} \frac{\poch{a} \poch{b} \poch{c}}{\poch{d} \poch{e} \poch{1}}x^n,
 \label{id3F2}\\
 F_4(a, b; c, d; x, y)=& \sum\limits_{m,n\ge0 } \frac{\left(a\right)_{m+n} \left( b \right)_{m+n}}{\left( c\right)_{m}
 \left(d\right)_n (1)_m (1_n)} x^m  y^n.
 \label{idF4}
\end{align}

\section*{Linear transformations}

\begin{align}
 {}_2F_1(a,b,c;1)=& \frac{\Gamma(c)\Gamma(c-a-b)}{\Gamma(c-a)\Gamma(c-b)} ,\label{id2}\\
 {}_2F_1(a,b,c;z)=& (1-z)^{(c-a-b)}{}_2F_1(c-b,c-a;c;z)\label{id1},\\
 {}_2F_1(a,b,c;z)=&\frac{\Gamma(c) \Gamma(c-a-b)}{\Gamma(c-a) \Gamma(c-b)} {}_2F_1(a, b ; a+b-c+1 ; 1-z) \qquad \qquad (|\arg (1-z)|<\pi)
 \label{id4}\\
&+(1-z)^{c-a-b} \frac{\Gamma(c) \Gamma(a+b-c)}{\Gamma(a) \Gamma(b)}  {}_2F_1(c-a, c-b ; c-a-b+1 ; 1-z), \nonumber
\\
{}_2F_1(a,b,c;z)=& (1-z)^{-a} \frac{\Gamma(c) \Gamma(b-a)}{\Gamma(b) \Gamma(c-a)} {}_2F_1\left(a, c-b ; a-b+1 ; \frac{1}{1-z}\right)
\qquad (|\arg (1-z)|<\pi)  \label{id3}\\
 &+(1-z)^{-b} \frac{\Gamma(c) \Gamma(a-b)}{\Gamma(a) \Gamma(c-b)} {}_2F_1\left(b, c-a ; b-a+1 ; \frac{1}{1-z}\right).
  \nonumber
\end{align}

\section*{Pochhammer identities}

\begin{equation}
\begin{aligned}
\pochm{0}_{0}=&1,\\
\pochm{0}_{m}=&0, \qquad  \qquad m\in \mathbb{N},\\
\pochm{0}_{-m}=&\frac{(-1)^m}{\pochm{1}_m}, \qquad m\in \mathbb{N},
\end{aligned}\qquad \qquad \qquad
\begin{aligned}
\pochm{a}_{-m}=&\frac{(-1)^m}{\pochm{1-a}_m},\\
\pochm{a}_{m+n}=&\pochm{a}_m \pochm{a+m}_n.
\label{poch-identities}
\end{aligned}
\end{equation}

\section{Macaulay2 example}
\label{glossary}

In this Appendix, we will give a short example using \textsl{Macaulay2} \cite{M2} of the algorithm to compute fake indicial ideal for 
the party hat integral. The matrix $\sfA$  associated with this integral  is given in Eq.\eqref{party-hat-matrix}. Our 
starting point will be to compute the toric ideal $I_\sfA$. Using the procedure in Part II (Toric 
Hilbert Schemes) of Ref.\cite{sturmfels:2001} gives

\begin{verbatim}
i2 : A={{1, 1, 1, 1, 1, 1}, {1, 1, 0, 0, 0, 0}, {0, 0, 1, 1, 0, 1}, {1, 0, 1,
                  0, 1, 1}, {0, 1, 0, 1, 1, 1}};
i3 : toricIdeal A
o3 = ideal(b*c - a*d).
\end{verbatim}

\noindent We write

\begin{align}
 I_\sfA=\braket{\partial_2\partial_3-\partial_1\partial_4}.
\end{align}

\noindent The next step it to compute  the initial ideal of $I_\sfA$ with respect to a weight vector $w$. Using the built in 
\textsl{Macaulay2} package $D$-modules to obtain the initial ideal with respect to
$w=(0,1,1,1,1,1)$ gives

\begin{verbatim}
i4 : loadPackage "Dmodules";
i5 : D=QQ[a,b,c,d,e,f,Da,Db,Dc,Dd,De,Df,
     WeylAlgebra=>{a=>Da,b=>Db,c=>Dc,d=>Dd,e=>De,f=>Df}];
i6 : IA=ideal(Db*Dc-Da*Dd);
o6 : Ideal of D
i7 : toString inw(IA,{0,-1,-1,-1,-1,-1,0,1,1,1,1,1})
o7 = ideal(Db*Dc)
\end{verbatim}

\noindent Notice that $\mathbb{K}[\partial_1,\dots,\partial_n]$ is a commutative ring and hence
$\text{in}_{(-w,w)}(I_A)=\text{in}_w(I_A)$. We  have the monomial ideal

\begin{align}
\text{in}_w(I_A)= \braket{\partial_2\partial_3}.
\end{align}

\noindent In \textsl{Macaulay2} the \texttt{standardPairs} function computes the Standard Pairs of 
a monomial ideal. We define a monomial ideal in the commutative ring $\mathbb{Q}[a,\dots,f]$ and compute 
its standard pairs.

\begin{verbatim}
i8 : R=QQ[vars(0..5)];
i9 : Iinw=monomialIdeal(b*c);
o9 : MonomialIdeal of R
i10 : standardPairs Iinw
o10 = {{1, {d, c, a, e, f}}, {1, {d, b, a, f, e}}}
o10 : List
\end{verbatim}
\begin{align}
\mathcal{S}( \text{in}_{(w)}(I_A))=& \{\{1, \{1,3,4,5,6\}\},\{1,\{1,2,4,5,6\}\}\} ,\\
\text{ind}_{w}(I_A)=& 
\bigcap\limits_{(\partial^a, K)\in \mathcal{S}( \text{in}_{w}(I_A))}
\braket{(\theta_j-a_j), j\notin K}= \braket{\theta_2}\cap\braket{\theta_3},
\end{align}
\noindent therefore

\begin{align}
\text{fin}_{w}(H_{\mathsf{A}}(\kappa))=& \braket{\theta_2 \theta_3}+\braket{A\theta-\kappa^T}.
\end{align}

\bibliographystyle{JHEP}

 \renewcommand\bibname{References} 
\ifdefined\phantomsection		
  \phantomsection  % makes hyperref recognize this section properly for pdf link
\else
\fi
\addcontentsline{toc}{section}{References}

\bibliography{./delacruz-AFeynman-data.bib}

\end{document}